\documentclass[11pt]{article}

\usepackage{graphicx}
\usepackage{amsmath}
\usepackage{amsfonts}
\usepackage{amsthm}
\usepackage{hyperref} 
\usepackage{bbm}
\usepackage{amsopn,amssymb,url,mathrsfs,a4wide}
\usepackage{tikz}
\usepackage{booktabs}
\usepackage[margin=2.25cm]{geometry}
\usepackage{fancyvrb}
\usepackage{listings}
\usepackage{float}
\usepackage{verbatim}
\usepackage{caption}
\usepackage{subcaption}
\usepackage{titlesec}
\usepackage{natbib}
\newcommand{\shp}{\xi}       			    

\newcommand{\site}{s}      		                    
\newcommand{\tm}{t}      		                    

\newcommand{\clim}{c}      		                    
\newcommand{\obs}{o}      		                    

\newcommand{\thresh}{u}      		                

\newcounter{ctr}

\newcounter{ctr1}

\newcounter{ctr2}

\newcounter{ctr3}

\newenvironment{theorem*}[1]{{\bf Theorem #1} \begin{itshape}}{\end{itshape}}

\newenvironment{corollary*}[1]{{\bf Corollary #1} \begin{itshape}}{\end{itshape}}

\newenvironment{proposition*}[1]{{\bf Proposition #1} \begin{itshape}}{\end{itshape}}

\newcommand{\ud}{\, {\rm d} \kern-.015em }

\newcommand{\modulus}[1]{\left| \kern.05em #1 \kern.05em \right|}
\newcommand{\norm}[1]{\left\| \kern.05em #1 \kern.05em \right\|}
\newcommand{\inner}[1]{\left\langle \kern.05em #1 \kern.05em \right\rangle }

\newcommand{\pick}[2]{\renewcommand{\arraystretch}{0.6}
\left( \kern-.4em \begin{array}{c} #1 \\ #2 \end{array} \kern-.4em \right) }

\newsavebox{\FVerbBox}

\titlespacing\chapter{0pt}{3pt plus 4pt minus 2pt}{3pt plus 2pt minus 2pt}
\titlespacing\section{0pt}{3pt plus 4pt minus 2pt}{3pt plus 2pt minus 2pt}
\titlespacing\subsection{0pt}{3pt plus 4pt minus 2pt}{3pt plus 2pt minus 2pt}
\titlespacing\subsubsection{0pt}{3pt plus 4pt minus 2pt}{3pt plus 2pt minus 2pt}
\titlespacing\paragraph{0pt}{3pt plus 2pt minus 2pt}{3pt plus 2pt minus 2pt}
\usepackage{amsmath}
\usepackage{authblk}
\usepackage{float}

\title{Inference for extreme spatial temperature events in a\\changing climate with application to Ireland}

\author[1]{Dáire Healy}
\author[2]{Jonathan Tawn}
\author[3]{Peter Thorne}
\author[1]{Andrew Parnell}

\affil[1]{Hamilton Institute, Maynooth University.}
\affil[2]{Department of Mathematics and Statistics, Lancaster University.}
\affil[2]{ICARUS, Department of Geography, Maynooth University.}
\date{}
\begin{document}
  \maketitle
  \begin{abstract}
\noindent We investigate the changing nature of the frequency, magnitude and spatial extent of extreme temperatures in Ireland from 1931 to 2022. We develop an extreme value model that captures spatial and temporal non-stationarity in extreme daily maximum temperature data. We model the tails of the marginal variables using the generalised Pareto distribution and the spatial dependence of extreme events by a semi-parametric Brown-Resnick $r$-generalised Pareto process, with parameters of each model allowed to change over time. We use weather station observations for modelling extreme events since data from climate models (not conditioned on observational data) can over-smooth these events and have trends determined by the specific climate model configuration. However, climate models do provide valuable information about the detailed physiography over Ireland and the associated climate response. We propose novel methods which exploit the climate model data to overcome issues linked to the sparse and biased sampling of the observations. Our analysis identifies a temporal change in the marginal behaviour of extreme temperature events over the study domain, which is much larger than the change in mean temperature levels over this time window. We illustrate how these characteristics result in increased spatial coverage of the events that exceed critical temperatures.
\end{abstract}

\vspace{0.3cm}
\noindent
Keywords: Climate change, generalised Pareto distribution, extreme values, heatwaves, missing data, non-stationarity, Pareto processes, spatial extremes, temperatures.
\vspace{0.5cm}

\section{Introduction}\label{SEC1}
\noindent The \citet[Chapter 11]{ipcc2021ch11} reports an observable change in extreme weather and climate events since around 1950. Characterisation of extreme temperature events is crucial for societal development, for estimating risks, and to enable the mitigation of their effects for many sectors, e.g., healthcare, economic growth, agricultural disruption, and infrastructure. \cite{Brown2008} observed a warming of both maximum and minimum temperatures since 1950 for most regions indicating an increasing number of warm days, longer heatwaves, and fewer cold extremes. 

In Ireland, changing extreme temperature behaviour has also been observed. \cite{Mcelwain2007} found that a warming of both maximum and minimum temperature observations occurred for all sites over 1961-2005. \cite{OSullivan2020} showed that the frequency of extreme temperature events for County Dublin has increased over the period 1981-2010. Both these approaches considered only the marginal behaviour of extremes. To the best of our knowledge, the only modelling of spatial extreme temperature events in Ireland is by \cite{Huser2020a}. They used a gridded Irish temperature data set, which has the potential to be over-smooth relative to the observed process, and fitted their model to these data under the assumption of stationarity over time and space. Under similar stationarity assumptions, \cite{Fuentes2013} and \cite{Cebrian2022} present analyses of extreme spatial temperatures for other locations.

We are interested in developing a model which captures the temporal evolution of spatial extreme temperature events over Ireland. This involves modelling how the marginal distributions vary over space, accounting for spatial dependence within extreme events, and modelling how these two elements vary over time. Our focus is on modelling extreme value data, however, for a spatial process, an extreme event can consist of abnormally high values in part of the region and typical values elsewhere \citep{Davison2012}.

Observational extreme event data are sparse and so they need to be used efficiently. The traditional statistical approach is to model these data with powerful probabilistic characterisations from extreme value theory. This theory provides a parsimonious asymptotic justification for extrapolation which enables us to describe the properties and behaviour of events which are more extreme than those previously observed. The theory by itself, however, will not provide information on how to spatially interpolate over heterogeneous geography or how to account for when the characteristics of complex spatial events change over time. Here we propose a novel approach to address these issues which exploits the physical knowledge of the climate processes from information given by fine-scale climate model data. We review existing extreme value methods for spatial and temporal processes and outline our strategies for using climate model data.

The theory of univariate extreme values for stationary processes \citep{Leadbetter1983}, and the associated statistical models \citep{Coles2001}, fully determine a simple parametric distributional family, namely the generalised Pareto distribution (GPD), as the non-degenerate limit distribution for the normalised excesses of a threshold, as that threshold tends to the upper endpoint of the marginal distribution. The GPD has been very widely used in diverse applications since the exposition of \citet{Davison1990}. To deal with non-stationarity, the GPD parameters have been allowed to change smoothly with covariates, initially using fully parametric regression models and more recently with a range of different non-parametric smoothing methods \citep{Chavez-Demoulin2005,Youngman2020}. We model the upper tails of the marginal distribution of the temperature process using the GPD, with covariates selected from space, time, information from climate models, and established measures/causes of climate change.

The most established approach to spatial extreme modelling uses max-stable process models \citep{DeHaan2006}. These processes are the class of non-degenerate limiting distributions of linearly normalised site-wise maxima, typically fitted to annual maxima data observed at each site in a set of locations over years. They are a natural extension of univariate block maximum limit theory, and so have generalised extreme value distributions for their margins. \cite{brown1977} introduced a widely used subclass of these models, derived from Gaussian random fields, known as Brown-Resnick processes, with \cite{Davis2013a} applying this model to spatio-temporal data.

The major problem with max-stable models is that they do not model, and so cannot capture, spatial patterns of observed extreme events. Inference using these models can lead to biased estimation of dependence \citep{Huser2020a}. A recent development in the modelling of spatial threshold exceedances is the generalised $r$-Pareto process \citep{Thibaud2015, DeFondeville2018a, DeFondeville2022}. Generalised $r$-Pareto processes, like max-stable processes, exhibit a strong form of dependence, known as asymptotic dependence (defined in Section~\ref{SEC4_2}) between all sites. This implies that for an event which is extreme at any location in space, there is a positive probability that this event will be extreme everywhere else in the spatial domain. For processes over spatial domains that are large relative to the scale of the spatial dependence of the process, this is an unrealistic assumption. More flexible spatial models, building from those in \cite{Wadsworth2019} are discussed in \citet[Section 7.1]{Healy2023sup}. 

In Section~\ref{SEC4} we define these extremal dependence properties precisely and provide evidence that generalised $r$-Pareto processes are suitable for daily maximum temperatures over Ireland. We identify extreme spatial fields, based on observations at $d$ sites, as those which exceed a sufficiently high threshold for a risk function $r:\mathbb{R}^d \to \mathbb{R}_+$. We model these fields as realisations of a Brown-Resnick Pareto process, which is closed under marginalisation \citep{Engelke2020}, an important property given the time-varying level of missing temperature data in our application.
 
We have observational daily maximum data for a network of 182 Irish temperature stations, with only $38\%$ of these having more than 30 years of data due to differential operational periods and quality controls, which is further compounded by spatial selection bias in the station locations. We also have a rich spatio-temporally complete data set generated from a climate model, giving daily maximum temperatures over 56 years on a fine grid over Ireland. These climate model data are not conditioned on the observed weather, so their values on any given day have no correlation to the observed data, but they have similar probability distributions to the observed data at the associated sites and time of the year. The climate model data have no missing values or location biases, they are on a dense regular grid and incorporate the impact of known geophysical structures on the temperature process.

Although it may be tempting to analyse the simpler climate model data than the observed station data, climate models involve some abstraction of the physical processes they model and so tend to under-predict extreme events in magnitude and to over-estimate dependence owing to the climate model's smoothness over space and time, see Sections~\ref{SEC3} and \ref{SEC4}. So direct analysis of the climate data is not ideal but clearly, they offer vital additional information to the observational data. Various attempts have been made to downscale the climate model data to produce a proxy for the observed data which gives a spatial and temporally complete data set, e.g., \cite{Maraun2017}, with the focus to date being on assessing marginal features. We prefer to let the observational data stand for themselves, particularly in relation to the information they provide about temporal non-stationarity as the climate model data have trends determined by the climate model configuration that is an imperfect representation of the real-world processes.

The novelty of our paper is achieved through the use of state-of-the-art extreme value methods for marginal distributions, spatial dependence and temporal non-stationarity which collectively exploit knowledge from climate science and through the use of appropriate metrics for describing changes in spatial extreme events. Our use of climate science relies heavily on how our inference for the observational temperature data leverages core information from the climate model data, i.e., parameter estimates (within sample quantiles and GPD parameters) over space, and through our careful assessment of, and sensitivity to, the effects of the inclusion of various climate-based covariates.

The paper is organised as follows. Section \ref{SEC2} details the observational and climate model data used. Sections~\ref{SEC3} and \ref{SEC4} describe the marginal and dependence modelling of the process respectively, in each case accounting for their changing behaviour over time. In Section~\ref{SEC5} we use the model to explore how the properties of spatial extreme events have changed over time. Conclusions and a broader discussion are given in Sections~\ref{SEC6} and \ref{SEC7}. All our code and instructions on how to access the data are available on GitHub\footnote{github.com/dairer/Extreme-Irish-Summer-Temperatures}.\\

\section{Data}\label{SEC2}
\noindent We start with a note on nomenclature since we use multiple data sets which differ in their structure and use. We use the term `station data' to refer to data taken directly from weather stations. These are irregularly located and suffer from missing values. The term `climate model data' refers to physics-based simulations of the weather system which are run on high-resolution grids and do not aim to match individual weather events, rather they model the spatio-temporal dynamics of the weather system. Finally, `observation-based data products' are gridded data sets which arise from some form of statistical or physical interpolation of station data. \\

\subsection{Station data}
\noindent Our daily maximum temperature data comprise 182 Irish temperature stations compiled from two sources, the locations of which are shown in Figure~\ref{fig:obs_map}. For the Republic of Ireland, data for 151 stations came from Met Éireann's archive\footnote{Copyright Met Éireann. Source: www.met.ie/climate/available-data/historical-data.}. Data for 31 Northern Ireland sites were obtained through the CEDA archive \citep{ukmo-midas}. Collectively these data have many missing values, with availability of data decreases further back in time. We have more than twice the data from the 1950s than the 1940s, with all stations (except one) pre-1950 being coastal. We have 56\% of daily values observed in the last 30 years, and only 0.53\% observed before 1942. No single day has data for every station. The average span of data for each station is about 30 years, with observations ranging from 1931 to 2022. The sites with the most data tend to be located near the coast reflecting historical and present-day observational priorities.

Our interest is in extreme warm temperatures in Ireland, so we restrict our analysis to data from the summer (JJA). This choice is supported by the finding that 93\% of all the days with temperatures above the 99\% site-wise marginal quantile occur in summer, with this proportion increasing with the level of marginal quantile. Furthermore, we assume that within each summer the process is temporally stationary, with exceedances being reasonably spread across the summer. Exploratory analysis supporting both of these choices is reported in \citet[Section 2.3]{Healy2023sup}.\\

\subsection{Climate model data}
\noindent Climate models are mathematical representations of the physical processes driving weather and climate and represent our best understanding of these natural phenomena \citep{Giorgi2019}. Climate models are broadly run on two scales; large-scale global climate models (GCMs) and finer-scale regional climate models (RCMs). An RCM is informed by the GCM at its boundary. Crucially, climate model data do not correlate with the observed time-evolution of weather, rather they have probabilistic structures that reflect plausible weather sequences which could occur. They are typically designed to investigate the effect of potential external forcing on the climate system by, e.g., increases in greenhouse gas concentrations arising from anthropogenic activities. We identify and exploit physical and topographical features in the output of these models and use them to adjust for spatial and temporal bias in the observed data set.

\begin{figure}[ht]
 \centering
 \includegraphics[width =12cm]{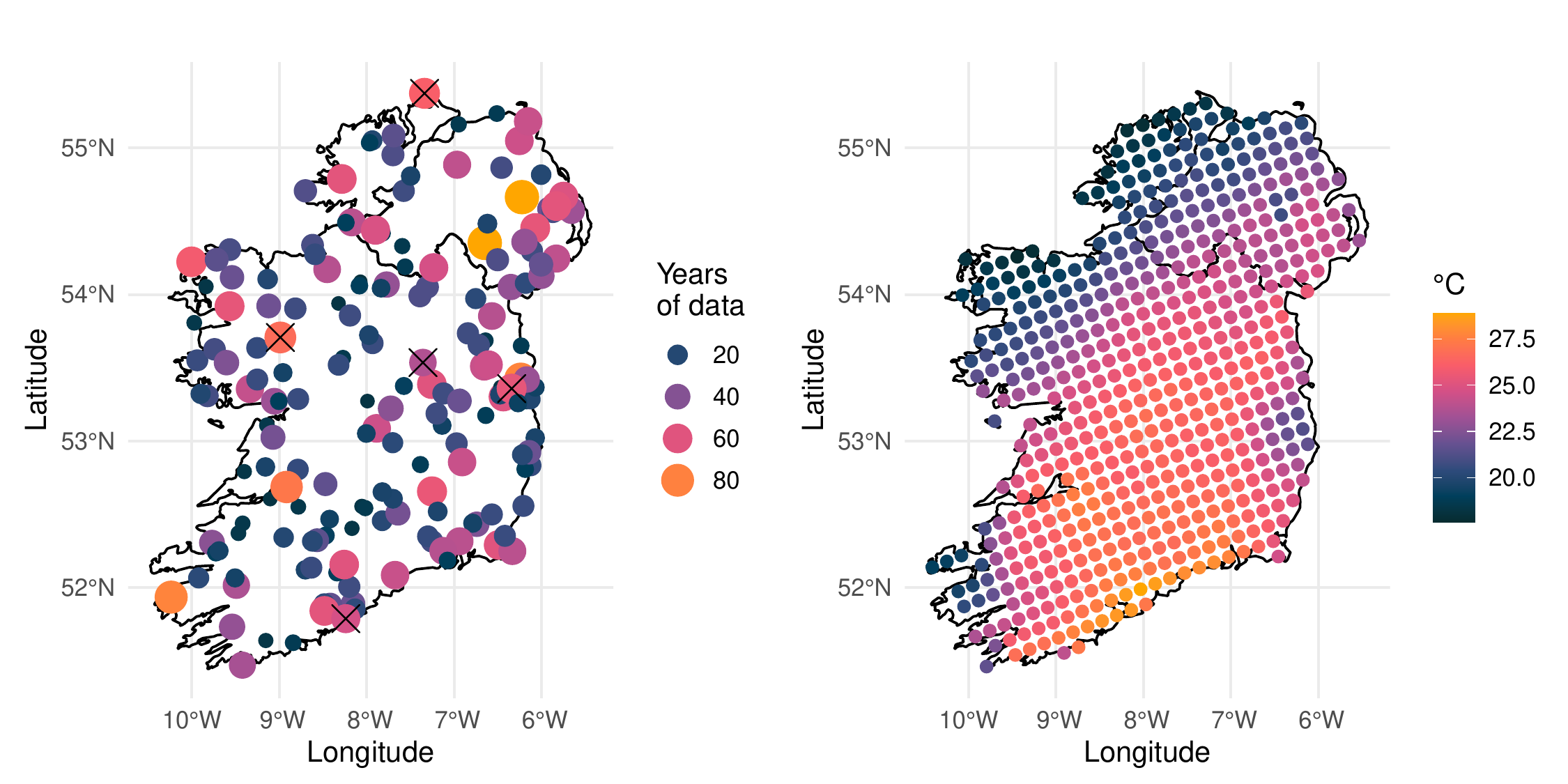}
 \caption{Ireland data locations: (left) station data sites, with the amount of data indicated by colour and size. Sites marked with an `X' correspond to Malin Head (North), Roches Point (South), Phoenix Park (East), Claremorris (West) and Mullingar (Centre); (right) climate model data from MOHC-HadREM3-GA7-05 showing a generated extreme temperature event.}
 \label{fig:obs_map}
\end{figure}

We use RCMs for their detailed topographical information and their physical description of temperature processes. When relying only on climate models to understand extreme weather it is common to consider several GCM/RCM combinations, each with different initial conditions and future climate scenarios. This choice will have limited impact for us as we use station data to describe the magnitude and frequency of temperature events, and the climate model data only to inform non-temporal features. We use data from the CLMcom-CLM-CCLM4-8-17 RCM combined with the ICHEC-EC-EARTH GCM. Specifically, we have daily maximum temperatures over a 56-year period (created using the atmospheric climate drivers from 1950 to 2005) on a regular grid of 558 points over Ireland (corresponding to a $0.11^2$ degree resolution). Figure~\ref{fig:obs_map} (right) shows the values for the day with the largest average temperature over Ireland in this data set. This plot illustrates two features which we exploit in Sections~\ref{SEC3} and \ref{SEC4} respectively. Firstly, the RCM provides much greater spatial coverage in the interior of Ireland than the stations in Figure~\ref{fig:obs_map} (left). Secondly, extreme temperature events can be very widely spread across Ireland, since even the sites with the lowest values on this day have temperatures in their marginal distribution upper tails.\\

\subsection{Covariates: Observation-based data products}\label{sec:hadcrut_details}
\noindent To model temporal non-stationarity of extreme temperature data it is common to use time as the sole covariate, although this will have severe limitations outside the range of the data as potential emission scenarios diverge. Instead, use the time-varying covariates that climate scientists believe best represent changes in observed mean temperatures. These are predictable into the future under different emission scenarios. We use two covariates, smoothed monthly average temperature anomalies for the global average $M^{G}_{t}$, and for the grid box over Ireland $M^{I}_{t}$, from the observation-based data product HadCRUT5, see \citet[Section 2.1]{Healy2023sup} for details and plots of the covariates. Over 1942-2022, both covariates increase by $\sim 1^{\circ}$C, with the change accelerating.

Our exploratory analysis, using spline-based models, identified that the shortest distance to the coast, for each site, was a potential descriptor of the change of temporal trends, we define this covariate by $C(\boldsymbol{\site})$, for each site $\boldsymbol{\site}$. We consider the covariate of the annual CO$_2$ emissions (CO$_{2,t}$) for Ireland, see plot an details in \citet[Section 2.2]{Healy2023sup}. Since there is strong collinearity present in the collective covariates $\boldsymbol{z}_t:=(t, M^{G}_{t}, M^{I}_t, \mbox{CO}_{2,t},C(\boldsymbol{\site}))$ we only use one of these at a time in each model.\\

\section{Marginal models}\label{SEC3}
\subsection{Overview and Strategy} \label{sec:overview}
\noindent Let $X_{\text{\obs}}(t,\boldsymbol{\site})$ denote the station data comprising summer maximum daily temperature at time $t$ and site $\boldsymbol{\site}$, and let $X_{\text{\clim}}(t,\boldsymbol{\site})$ be the equivalent process from the climate model data. We assume temporal stationarity within each year for each site and each process. Here $t \in \mathbb{N}$ indexes summer days within and across years and $\boldsymbol{\site} \in \mathcal{S} \subset \mathbb{R}^2$, where $\mathcal{S}$ denotes Ireland, with $\boldsymbol{\site}$ corresponding to the vector of latitude and longitude. We have data on the two processes at $\mathcal{S}_o\subset \mathcal{S}$ and $\mathcal{S}_c\subset \mathcal{S}$ and at times $\mathcal{T}_o$ and $\mathcal{T}_c$ respectively. For $\mathcal{T}_o$ we also have missing data for some of the stations as discussed in Section~\ref{SEC2}. Throughout we use the subscripts to identify the type of process, though the indexing is dropped when discussing methods which apply similarly to both processes.

In Section~\ref{sec:below} we propose a spatial and temporal quantile regression model for the data to derive an estimate of the distribution function of $X(t,\boldsymbol{\site})$. As the tails of this distribution are particularly important to model well we introduce a threshold $u(\boldsymbol{\site})$, which is fixed over time but varies in space, above which we replace the quantile model with the generalised Pareto distribution (GPD) parametric model with temporal and spatial covariates. The justification for our choice of a constant threshold over time is discussed in Section~\ref{const_thresh_discussion}. Novelty in our approach comes from using estimates from the $X_{\text{\clim}}$ process to infer features of the $X_{\text{\obs}}$ process, which is appealing as $|\mathcal{S}_o| \ll |\mathcal{S}_c|$ and that $|\mathcal{T}_o| \ll |\mathcal{T}_c|$ for most sites. Given all these considerations, we need to estimate thresholds, the temporally varying marginal distributions over $\mathcal{S}$ for below the thresholds, the GPD parameters for above the thresholds, and to do this for both the $X_{\text{\clim}}$ and $X_{\text{\obs}}$ processes.

When analysing the climate model data we need to account for the following issues. First, our use of these data is to improve our spatial mapping and to overcome issues of missing data in the analysis of the observational data. Second, climate model data can show different time dynamics from that of the observed process since they are based on incomplete physics and forcing detail. We want our analysis to be robust to temporal non-stationary aspects of the climate model data, so we assume that $X_c(t,\boldsymbol{\site})$ are temporally stationary in our analyses. As the trend in the climate model data is $4\%$ of the variation in the data at each site, this is not too restrictive an assumption.\\

\subsection{Modelling the body of the distribution}
\label{sec:below}
\noindent Given the issues raised in Section~\ref{sec:overview} about $X_c(t,\boldsymbol{\site})$, we take the following simple approach for the inference of its distribution. For site $\boldsymbol{\site}$ we estimate the $\tau$th quantile of $X_c$ by using the empirical sample quantile for the climate model data at that site alone; we denote this estimator by $q_c^{(\tau)}(\boldsymbol{\site})$. We use this approach for all $\tau$ over the range $[0.01,0.99]$. This estimator is reliable as we have sufficient data (5152 days with none missing) and, due to the climate model data being numerical model output, their spatial variation is very smooth, so statistical spatial smoothing methods are more likely to induce bias than to enhance the analysis through information sharing. 

For the analysis of $X_{\text{\obs}}(t,\boldsymbol{\site})$ the issues of spatial sparsity of stations, limited data, varying periods of records of stations, and the need to account for temporal variations, lead to a different approach than for $X_{\text{\clim}}(t,\boldsymbol{\site})$. We follow the approach of \citet{Yu2001} and that of the R package \texttt{evgam} \citep{Youngman2020} by using the asymmetric Laplacian distribution (ALD) for quantile regression to estimate a range of spatially and temporally varying $\tau$th quantiles, $q_o^{(\tau)}(t,\boldsymbol{\site})$, for a grid of $\tau\in [0.01,0.99]$, for all $t \in \mathcal{T}$ and $\boldsymbol{\site} \in \mathcal{S}$. The density function of the ALD$_\tau$ is
\begin{align}\label{eq:ald}
 f_{\mathrm{ALD}_\tau}(y ; q, \psi)=\tau(1-\tau)\psi^{-1} \exp \left\{-\rho_{\tau}\left({y-q}\right)\psi^{-1}\right\},\; y \in \mathbb{R}, 
\end{align}
where $\rho_{\tau}(z)=\{\tau-I(z<0)\}z$ is the check function, $q\in \mathbb{R}$ is a location parameter, corresponding to the $\tau$th quantile of interest, and $\psi>0$ is a scale parameter. We assume that $q$ and $\psi$ vary smoothly over $\mathcal{S}_o$ and $\mathcal{T}_o$.

For estimating $q_o^{(\tau)}(t,\boldsymbol{\site})$ and $\log \{\psi_{o}^{(\tau)}\}$ we consider not just $t$ and $\boldsymbol{\site}$ as covariates but also incorporate as potential covariates the associated quantile from the climate model data $q_c^{(\tau)}(\boldsymbol{\site})$ and each of the climate-based covariates of Section~\ref{sec:hadcrut_details}. The former provides richer spatial information that is not captured in the observational data set, and the latter gives a causal set of time-varying covariates. Details of the analysis using these models are presented in Section~\ref{sec:DA}. 

To provide estimates for all $\tau$, we fit this model separately for a grid of $\tau$ values and use a cubic interpolation spline for each $\boldsymbol{\site}$ to give a continuous estimate over $0.01\le \tau \le 0.99$ for $\hat{q}_o^{(\tau)}(t,\boldsymbol{\site})$. We kept the grid of $\tau$ values relatively coarse to avoid issues of quantile estimates crossing. This gives us an estimate of the distribution function of $X_o(t,\boldsymbol{\site})$ as 
\begin{equation}
F_{\text{\obs},t,\boldsymbol{\site}}(\hat{q}_o^{(\tau)}(t,\boldsymbol{\site})):=
\Pr\{X_o(t,\boldsymbol{\site})<\hat{q}_o^{(\tau)}(t,\boldsymbol{\site})\}=\tau \mbox{ for all }\tau.
\label{eq:distfunct}
\end{equation}
This model provides estimates for all quantiles for any $\boldsymbol{\site} \in \mathcal{S}_c$, not just $\mathcal{S}_o$, and at all times where we have the covariates, e.g., not just for $t \in \mathcal{T}_o$. At each site $\boldsymbol{\site}$, below the threshold $u(\boldsymbol{\site})$ (defined in Section~\ref{sec:above}) we use this distributional model $F_{\text{\obs},t,\boldsymbol{\site}}$.\\

\subsection{Modelling the tails of the distribution}
\label{sec:above}
\noindent It is well known that quantile regression, and hence the ALD model, is unreliable for estimating quantiles in the tails of the distribution and provides no means to extrapolate beyond the observed data. As the upper extremes of the distribution of $X_o(t,\boldsymbol{\site})$ are important to us we chose to use a different model based on extreme value methods. This enabled us to produce a model that is continuous over all $t$ and $\boldsymbol{\site}$, with the extreme value model being used above a high threshold, and the ALD model describing data below.

One option is to have a threshold, $u(t,\boldsymbol{\site})$, that varies over time and space computed for a given quantile, e.g., $u(t,\boldsymbol{\site})=\hat{q}^{(\tau)}(t,\boldsymbol{\site})$, for a choice of $\tau$. However, in extreme value inference, it is well-known that it is difficult to objectively select a threshold or to account for the uncertainty in that choice \citep{Northrop2017a}. Here it is the temporal change in extreme events which is of most interest, and this trend is small relative to other sources of variations in the data. We do not select a time-varying threshold using information from the body of the distribution, as this may bias results for the extremes. Instead, we choose the threshold to be constant over time but varying over space, i.e., $u(\boldsymbol{\site})$. Our choice is discussed further in Section~\ref{SEC7}.

To reduce subjectivity, for each site and for both $X_c$ and $X_o$ processes we use a common exceedance probability for the fixed-over-time threshold. Based on the use of standard extreme value threshold selection methods for stationary processes \citep{Coles2001} which we applied at each site/process separately, we identified that the 90\% quantile was suitable. 
For the reasons discussed in Sections~\ref{sec:overview}
and \ref{sec:below} we use the site-specific 
90\% empirical sample quantile for $u_c(\boldsymbol{\site})$ 
but a model-based estimate for $u_o(\boldsymbol{\site})$. 
Specifically, we fit the model for density~\eqref{eq:ald} with $\tau=0.9$
with the location parameter structured as 
$u_o(\boldsymbol{\site}):=
 q_o^{(0.9)}(\boldsymbol{s})= \beta_0 + \beta_1 u_c(\boldsymbol{\site})$,
with $(\beta_0,\beta_1)$ parameters. Thus the climate model data provides a means by which the spatially varying threshold $u_o$ for the observed data $X_o$ can be estimated. This routine aims to overcome the data quality limitations and to provide estimates for all $\boldsymbol{s} \in \mathcal{S}_c$.

For a given threshold there are two remaining elements required to model the extremes, namely the threshold exceedance probability $\lambda_o(t,\boldsymbol{s})$ and the distribution $H$ of the excesses of the threshold
\citep{Chavez-Demoulin2005}. We consider these in turn. We estimate 
$\lambda_o(t,\boldsymbol{s})$ from the model for the body of the distribution, using the set of estimated distribution functions~\eqref{eq:distfunct}. Specifically, 
\[
\lambda_o(t,\boldsymbol{s})=1-\tau_{u_o}(t,\boldsymbol{s}),
\]
where $\tau_{u_o}(t,\boldsymbol{s})$ is the value of $\tau$, at time $t$, which makes
$\hat{q}_o^{(\tau)}(t,\boldsymbol{\site})=u_o(\boldsymbol{\site})$. If there is no temporal non-stationarity in $X_o(t, \boldsymbol{s})$ then by construction of the threshold $\thresh_o(\boldsymbol{s})$, if that was correct, we would have $\lambda_o(t,\boldsymbol{s})=1-\tau=0.1$ across 
$t\in \mathcal{T}_o$ and $\boldsymbol{s}\in \mathcal{S}_o$. 

For each site $\boldsymbol{\site}$ we assume that excesses of the threshold $u(\boldsymbol{\site})$ follow a generalised Pareto Distribution (GPD); see \citet{Pickands1975} and \citet{Davison1990} for the probabilistic justification and properties. The GPD$(\sigma,\xi)$ has distribution function
\begin{equation}
 H(y; \sigma,\shp)=
	1 - (1 + \shp y / \sigma )_{+}^{-1/\shp} 
\end{equation}
for $y>0$, with a shape parameter $\shp \in \mathbb{R}$ and a scale parameter $\sigma>0$, with the notation $x_+=\max(x,0)$, and $\xi=0$ is obtained by taking the limit as $\xi\rightarrow 0$. When $X(t, \boldsymbol{s}) > \thresh(\boldsymbol{s})$ the threshold excess, $Y(t, \boldsymbol{s})= X(t, \boldsymbol{s}) - \thresh(\boldsymbol{s})$, is taken to be distributed 
\begin{align}\label{eq:gpd}
Y(t, \boldsymbol{\site}) \sim \text{GPD}(\sigma(t, 
\boldsymbol{s}), \shp),
\end{align}
where we discuss the choice of models for $\sigma(t, \boldsymbol{s})$ below, and where the shape parameter is taken to be constant over time and space. This choice of homogeneity for the shape parameter for both $X_o$ and $X_c$ (i.e., values $\xi_o$ and $\xi_c$ respectively) is supported by exploratory analysis in \citet[Section 3.2]{Healy2023sup}, but it is 
typical in GPD modelling as there is limited evidence against this in almost all applications, and for the pragmatic reason that even a homogeneous value is difficult to estimate well. Furthermore, this choice reduces the risk of parameter identifiability problems \citep{Davison2012}. 

Combining $\lambda_o(t,\boldsymbol{s})$ with the model for $H$ gives our overall marginal distributional model $F_{o; t, \boldsymbol{s}}$ for the upper tail of $X_o(t, \boldsymbol{s})$. Specifically for $y> \thresh_o(\boldsymbol{s})$ we have
\begin{equation}\label{eq:tailF}
F_{o; t, \boldsymbol{s}}(y) = 1-\lambda_o(t,\boldsymbol{s})[1-H(y-u_o(\boldsymbol{s}); \sigma_o(t, \boldsymbol{s}),\shp_o)].
\end{equation}
As with the estimation of the quantiles below $u_o(\boldsymbol{\site})$ for $X_o$, we use information from $X_c$ to provide a spatial covariate for $\sigma_o(t, \boldsymbol{s})$. We aim to learn about temporal non-stationarity exclusively from the observational data, so only information about the spatial variation of the marginal tail distribution is taken from $X_c$. We fit a model of the form
\begin{equation}\label{eq:clim_gpd}
 Y_{\text{\clim}}(t, \boldsymbol{\site}) \sim \text{GPD}\{\sigma_{\text{\clim}}( \boldsymbol{\site}), 
 \shp_{\text{\clim}}\},
\end{equation}
for the excesses of $u_c(\boldsymbol{\site})=q_c^{(0.9)}(\boldsymbol{\site})$. When modelling the climate model data we believe we have sufficient observations and spatial consistency, from their generation, to treat $\sigma_{\text{\clim}}( \boldsymbol{\site})$ as site-specific, i.e., 
not imposing any spatial smoothness on the GPD scale parameters over $\boldsymbol{\site}\in \mathcal{S}$. Clearly, it would be wrong to smooth well-estimated parameters spatially if we want to capture the relevant geophysical features of the climate system. As discussed in Section~\ref{sec:overview} we do not allow the temporal variation in the climate model to be informative about the observational data so we keep $\sigma_{\text{\clim}}$ constant over $t$.

Full likelihood inference is not possible as any realistic model for the station data is likely to be highly complex, requiring spatial and temporal dependence of the data to be modelled. Instead, we use a pseudo-log-likelihood
\begin{align}
   p\ell_c\left(\sigma_{c}(\boldsymbol{s}): \boldsymbol{s} \in \mathcal{S}_c; \shp_c\right) = 
   \sum_{\boldsymbol{s}\in \mathcal{S}_c}
   \left[\sum_{t\in \mathcal{T}_c} 
   \log h(y_{c,t,\boldsymbol{s}};\sigma_{c}(\boldsymbol{s}),\xi_{c}) \right],
\end{align}
constructed under the false assumption of spatial and temporal independence, with $h$ being the density function of the GPD. This inference approach is commonly used, e.g., by \citet{Davison2012}.
The maximisation of this function can be broken down into a series of 1-dimensional optimisations by alternating the maximisation over $\xi_c$ with the scale parameters fixed, and then exploiting the partition of the function $p\ell_c$ with respect to 
$\boldsymbol{s}$ when maximising over each $\sigma_{c}(\boldsymbol{s})$ in turn whilst treating $\xi_c$ as constant. Iterating in this way until convergence is achieved gives estimated values $\{\hat{\sigma}_{\text{\clim}}( \boldsymbol{\site}\}; \boldsymbol{\site}\in \mathcal{S}_c\}$ and $\hat{\xi}_c$.

Next, we model the extreme observational data excess above the threshold, 
$u_{o}(\boldsymbol{s})$,
denoted by $Y_{\text{\obs}}(\tm, \boldsymbol{\site})$. The generic form of each of the models we consider is
\begin{align}
 Y_{\text{\obs}}(t, \boldsymbol{\site}) &\sim \text{GPD}(\sigma_{\text{\obs}}\{\boldsymbol{z}_t, \hat{\sigma}_c(\boldsymbol{\site})\}, \shp_{\text{\obs}}),
\end{align}
 where we model $\log \sigma_{\text{\obs}}\{\boldsymbol{z}_t, \hat{\sigma}_c(\boldsymbol{\site})\}$ as either a parametric linear model of $\log\hat{\sigma}_c(\boldsymbol{\site})$ and the covariates $\boldsymbol{z}_t$ (defined in Section~\ref{sec:hadcrut_details}) or via a GAMs formulation. We denote the parameters of $\sigma_{\text{\obs}}$ by $\boldsymbol{\theta}$.
As with the inference for the climate model data we have to use a pseudo-log-likelihood, constructed under the false assumption of spatial and temporal independence. For the fully parametric model, the pseudo-log-likelihood is
\begin{align}
   p\ell_o\left(\boldsymbol{\theta}; \shp_c\right) = 
   \sum_{\boldsymbol{s}\in \mathcal{S}_o}
   \left[\sum_{t\in \mathcal{T}_o} 
   \log h(y_{o,t,\boldsymbol{s}};\sigma_{o},\xi_{o}) \right],
   \label{eqn:plike_GPD}
\end{align}
whereas in the GAMs setting $p\ell_o$ is adapted by incorporating an additive spline smoothing penalty term 
\citep{wood2006}. Given the use of a pseudo-penalised likelihood, we cannot use standard methods for the evaluation of parameter uncertainty and model selection. Instead, the approaches we use are discussed in Section~\ref{sec:Uncertain}, with our marginal tail inference for the data being presented in Section~\ref{sec:DA}. \\

\subsection{Model uncertainty quantification and selection}
\label{sec:Uncertain}
\noindent In cases where a pseudo-likelihood is used, as in Section~\ref{sec:above}, the most widely adopted method for model selection is to adapt standard information criteria to account for model/likelihood mis-specification to greater penalise complexity relative to a better pseudo-likelihood fit. For spatial extremes, the composite likelihood information criterion \citep[CLIC,][]{Davison2012} is used, which includes a first-order asymptotically motivated additive adjustment factor. Despite being used in many pseudo-likelihood approaches, we have chosen not to use CLIC for model selection. This is because the likelihoods for extremes are far from the asymptotic elliptical forms around the mode, yet CLIC relies on such asymptotic theory; CLIC measures only goodness of fit in the sample yet we have rich enough data to exploit out-of-sample model assessment; and for determining the parameter uncertainty we are not relying on asymptotic theory. Below we outline the alternative approaches we use.\\

\subsubsection{Bootstrap methods}
\noindent For both model selection and parameter uncertainty evaluation we generate bootstrapped samples $X_o^*$ of $\{X_{\text{\obs}}(t,\boldsymbol{\site}): t\in \mathcal{T}_o, \boldsymbol{\site} \in \mathcal{S}_o\}$ for a given marginal distribution model. These bootstrap samples need to preserve all spatial dependencies, short-range temporal dependence consistent with the passage of weather systems, missing data patterns of the observational data, and to exhibit the temporal non-stationarity of the fitted model.

For a given marginal model, the bootstrap takes the set of transformed observed data $\{x^U_{\text{\obs}}(t,\boldsymbol{\site})=F_{\text{\obs},t,\boldsymbol{\site}}(x_{\text{\obs}}(t,\boldsymbol{\site})): t\in \mathcal{T}_o, \boldsymbol{\site} \in \mathcal{S}_o\}$ where $F_{\text{\obs},t,\boldsymbol{\site}}$
is given by the two model components of 
Sections~\ref{sec:below} and \ref{sec:above}. The $x^U_{\text{\obs}}(t,\boldsymbol{\site})$ values are realisations of uniform$(0,1)$ random variables that are identically distributed over time for each $\boldsymbol{\site} \in \mathcal{S}_o$, but with the temporal and spatial dependence structure of the $X_o(t,\boldsymbol{\site})$ process retained. To these data, we apply a vector temporal block bootstrap, with details of block structure and adaptions to account for the missing data described in \citet[Section 3.4]{Healy2023sup}. For each bootstrapped data set $X^{U*}=\{X^{U*}_{\text{\obs}}(t,\boldsymbol{\site}): t\in \mathcal{T}_o, \boldsymbol{\site} \in \mathcal{S}_o\}$ 
we use the inverse of the distribution function $F_{o,t, \boldsymbol{s}}$ to create
the bootstrapped sample $X_o^*$ with elements
\begin{equation}
 X_o^*(t, \boldsymbol{s}) = F^{-1}_{o,t, \boldsymbol{s}}\{X^{U*}_o(t, \boldsymbol{s})\}.
\end{equation}
Applying this raw bootstrap method induced bias in parameter estimates, and hence in sampling distribution estimates. The bias stems from ties in the extreme bootstrapped data that this method produces. As the very largest observations in a dataset are known to be the most influential on the GPD model fit \citep{Davison1990} this is particularly problematic. There is negative bias in the estimate of the shape parameter of the GPD. Since the shape and scale parameters are negatively correlated, there is also a positive bias in the scale parameter estimator. To adjust for these biases we use a bootstrap error correction, as described in \citet[Section 3.5]{Healy2023sup}. This is a two-step procedure with a location adjustment to the bootstrapped shape parameter estimate, then the scale parameter is re-estimated after fixing the adjusted shape parameter.\\

\subsubsection{Cross validation}\label{sec:cv}
\noindent For model fit diagnostics, 
we use two types of cross-validation (CV) to evaluate the performance of our models on out-of-sample data \citep[Ch 7.]{Hastie2008}. We use standard 90-fold CV (90-CV) so that the data are divided into 90 groups (folds), where each fold is removed in turn and the model is fitted to the remaining folds. Since standard CV can perform poorly when the data have spatial or temporal correlation \citep{Roberts2017}, we also use a spatio-temporal CV (ST-CV) with 90 folds, corresponding to 30 spatial clusters of station data (i.e., divided spatially into 30 contiguous groups) and 3 temporal folds. Each temporal fold consists of every third week in the summer months, preserving long-term temporal non-stationarity. We define the 90 ST-CV folds as all combinations of spatial and temporal clusters, taking the intersection as a fold.

For each left out fold, we compute two different goodness-of-fit measures to evaluate out-of-sample performance, the root mean square error (RMSE) and the continuous ranked probability score \citep[CRPS,][]{Gneiting2014}. The RMSE evaluates the general closeness between the empirically estimated and predicted quantiles, whilst the CRPS aims to match both the calibration and the sharpness of these extremes quantiles \citep{Zamo2018}. Here the empirical quantile, $x^{(\tau)}_o(t, \boldsymbol{s})$, is evaluated using the ordered data at site $\boldsymbol{s}$ and the year which contains time $t$, whereas the predicted quantiles are estimated as $\hat{x}^{(\tau)}_o(t, \boldsymbol{s})= F^{-1}_{\text{\obs},t,\boldsymbol{\site}}(\tau)$ for quantile $\tau$ from the appropriate model. The comparisons between ${x}^{(\tau)}_o$ and $\hat{x}^{(\tau)}_o$, for the same $t$, $\boldsymbol{\site}$, and $\tau$, are averaged across the folds. Lower values of RMSE and CPRS generally indicate a superior fit.\\

\subsection{Marginal Data Analysis}
\label{sec:DA}

\subsubsection{Body of distribution}
\noindent Following exploratory analysis, we identified three potential models
for the body of the distribution, which we present in Table~\ref{tab:quant_reg_models} along with their CV RMSE. The first model serves as a base, in which the location parameter is constant over space and time for each quantile $\tau$. In the second model we allow the quantile regression to vary spatially by using the corresponding climate model data quantiles 
$q^{(\tau)}_{\mathrm{c}}(\boldsymbol{s})$ as a covariate. 
The third model also includes the temporal Irish mean temperature covariate $M^I(t)$. The inclusion of the climate model covariate reduces the RMSE for both types of CV, whereas $M^I(t)$ improves the CV scores further, though not as much. We fitted a number of other covariate combinations for $\boldsymbol{z}_t$, as well as using the principal components of $\boldsymbol{z}_t$ to avoid issues of collinearity. Overall we found the third model provides the best balance of simplicity and fit, so use this for subsequent analysis.

\begin{table}
\caption{\label{tab:quant_reg_models} Cross validation (RMSE) on the quantile regression analysis for the body of the distribution.}
 \centering
 \begin{tabular}{lcc}
 \hline
 \textbf{Model structure for $\hat{q}_{o}^{(\tau)}(t, \boldsymbol{s})$} & {\textbf{ST-CV}} & {\textbf{90-CV}}\\
 \hline
 $ \beta^{(\tau)}_0 $ & 1.456 & 1.454\\
 $ \beta^{(\tau)}_0 + \beta^{(\tau)}_1 q^{(\tau)}_{\mathrm{c}}(\boldsymbol{s}) $ & 1.336 & 1.347\\
 $ \beta^{(\tau)}_0 + \beta^{(\tau)}_1 q^{(\tau)}_{\mathrm{c}}(\boldsymbol{s}) + \beta^{(\tau)}_2 M^{\text{I}}(t)$ & \textbf{ 1.304} & \textbf{1.318}\\\hline
 \end{tabular}
\end{table}

\citet[Section 3.1]{Healy2023sup} provides estimates of $\beta_2^{(\tau)}$, which show a slight decrease with $\tau$ although the confidence intervals widen. For all $\tau$, $\beta_2^{(\tau)}=1$ appears consistent with the data, indicating that mean summer temperatures in Ireland are a good representation of the temporal change for all the body of the distribution. The estimates of $\beta_1^{(\tau)}$ (not plotted) decrease, approximately linearly, from $0.75$ to around $0.65$ with $0<\tau<1$, showing that the climate model is not giving identical descriptions to the station data, as the estimates differ from $1$ significantly and change with $\tau$.\\

\subsubsection{Tails of the distribution}
\begin{figure}[ht]
 \centering
 \includegraphics[width = 15cm]{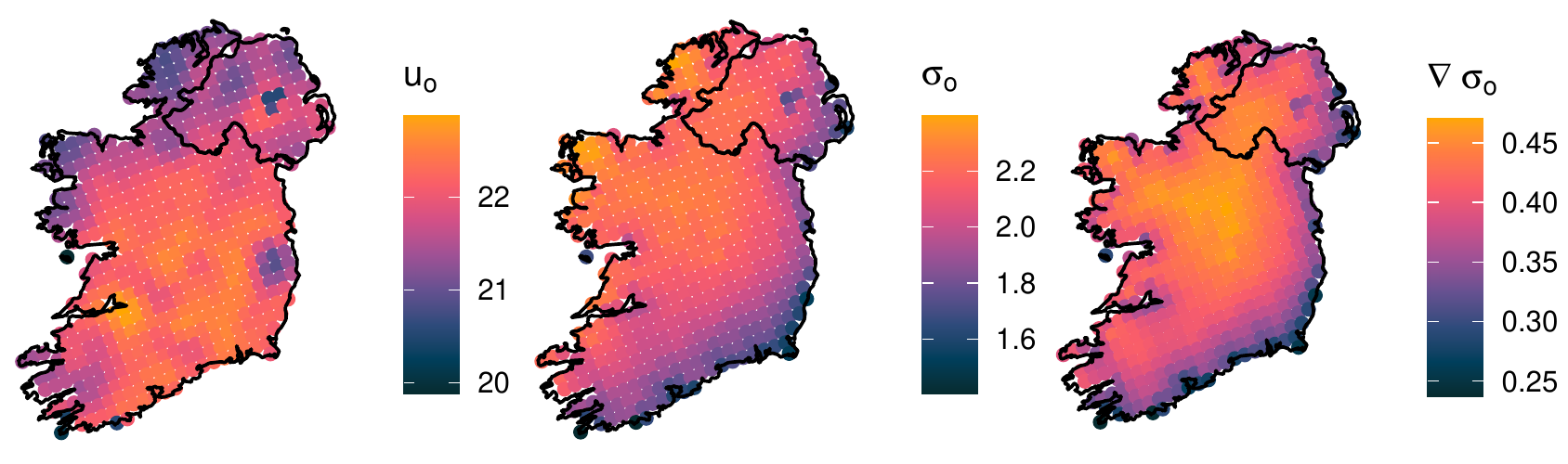}
 \caption{Estimated values of threshold $u_o(\boldsymbol{\site})$ (left), GPD scale parameter according to $M_2$ in $2020$, (centre), and the estimated change in the scale parameter since $1942$, $\nabla \sigma_\obs(\boldsymbol{s})$ (right).}
 \label{fig:thresh_scale}
\end{figure}

\noindent For selecting the threshold $u_o(\boldsymbol{s})$ we use the second model in Table~\ref{tab:quant_reg_models} with $\tau=0.9$,
providing a threshold that varies in space but not time. 
Figure~\ref{fig:thresh_scale} (left) shows the threshold $u_o(\boldsymbol{\site})$ over Ireland, with cooler temperature values on the west of Ireland and coastal regions on the south and north coasts. For this $u_o(\boldsymbol{s})$ we estimate the threshold exceedance probability $\lambda_o(t, \boldsymbol{s})$ and its spatial average $\lambda_o(t)=\int_{\boldsymbol{s}\in \mathcal{S}_c}
\lambda_o(t, \boldsymbol{s}) d\boldsymbol{s}/
|\mathcal{S}_c|$. Estimates of the latter are shown in \cite[Figure 4]{Healy2023sup}. The $\lambda_o(t)$ estimates show an increasing exceedance rate, with the average rate over time of $0.1$ reflecting the choice of the threshold, increasing by around 35\% with 95\% confidence interval 28-44\% from 1942-2020. We see the same features at individual sites, but with wider confidence intervals.

Table~\ref{tab:gpd_models_and_cv_res} presents a subset of the models that we explored for the GPD scale parameter: constant over space and time in model $M_0$; incorporating climate model data via $\sigma_{\text{\clim}}$, defined by expression~\eqref{eq:clim_gpd}, in $M_1$; allowing also for temporal non-stationarity via $M^I(t)$ in $M_2$ as both a constant rate of change and as an interaction with coastal distance $C(\boldsymbol{s})$. Other models were attempted with differing covariates and spline structure included but these failed to improve over models $M_0 - M_2$, however over a range of spline models we noticed that they were consistently suggesting evidence for different temporal trends on the coast relative to inland, hence our introduction of the $C(\boldsymbol{s})$ covariate. 

\begin{table}
\caption{\label{tab:gpd_models_and_cv_res}Models $M_0-M_2$ for GPD log-scale parameter, $\log \sigma_o(t,\boldsymbol{s})$ along with cross validation results and estimated shape parameter, $\xi_o$, with bootstrapped 95\% confidence intervals. Numbers in bold font show the lowest CV values.}
 \centering
 \begin{tabular}{lp{5cm}ccccc}
  & & \multicolumn{2}{c}{\textbf{ST-CV}} & \multicolumn{2}{c}{\textbf{90-CV}}& $\xi_o$ 
  \\
 & & RMSE & CRPS & RMSE & CRPS &\\ 
 \hline
 $M_0$ & $ \beta_0 + \beta_1 \ln \sigma_c(\boldsymbol{s}) $ & 0.953 & 0.905 & 0.928 & 0.882 & -0.152 (-0.237, -0.092)\\
 $M_1$ & $\beta_0 + \beta_1 \ln \sigma_c(\boldsymbol{s}) + \beta_2 \text{M}^I(t)$ & 0.946 & 0.904 & 0.917 & 0.880 & -0.156 (-0.204, -0.110) \\
 $M_2$ &$ \beta_0 + \beta_1 \sigma_c(\boldsymbol{s})+ \beta_2 \ln\text{C}(\boldsymbol{s}) + \newline \beta_3 \text{M}^I(t) + \beta_4 \ln\text{C}(\boldsymbol{s}) \text{M}^I(t)$ & \textbf{0.938} & \textbf{0.902} & \textbf{0.907} & \textbf{0.878} & -0.158 (-0.194, -0.110)\\ 
 \hline
 \end{tabular}
\end{table}

Table~\ref{tab:gpd_models_and_cv_res} presents our model selection diagnostics based on CV metrics (CV-CRPS and RMSE). 
All four approaches favour model $M_2$, with $M_0$ and $M_1$ having similar, slightly inferior performance, and we find that $M_0$ is too simplistic. Models $M_0-M_2$ estimate the coefficient of $\log \sigma_c$ as close to $1$ in all cases, showing that the climate model is providing very helpful information as a spatial covariate. The estimates of the shape parameter $\xi_o$ are also given in Table~\ref{tab:gpd_models_and_cv_res}. As the GPD scale parameter model is made increasingly flexible (from model $M_0$ to $M_2$) the value of $\xi_o$ decreases, lightening the tail decay, indicating that each model is progressively reducing sources of variation in the tail.
Since there is some uncertainty in the marginal model choice we take $M_0$, $M_1$ and $M_2$ through the spatial dependence analysis to assess the sensitivity of the risk measures, with details for model $M_2$ reported here and for $M_0$ and $M_1$ in \citet{Healy2023sup}.

Model $M_2$ shows that the most variable excess distribution is on the west coast (see centre plot in Figure \ref{fig:thresh_scale}), with a decay in values from west to east, so almost the opposite of the behaviour of $u_o(\boldsymbol{s})$. We also investigated the estimated change in the scale parameter over the observation period, denoted 
$\nabla \sigma_\obs(\boldsymbol{s})= \sigma_\obs(2020, \boldsymbol{s})-\sigma_\obs(1942,\boldsymbol{s})$, see Figure \ref{fig:thresh_scale} (right),
and found it to be largest in the centre of Ireland, with the change there being close to double that on the coast. Everywhere the scale parameter is increasing over time, leading to warmer extreme temperatures. 

\begin{figure}[ht]
 \centering
 \includegraphics[width = 10cm]{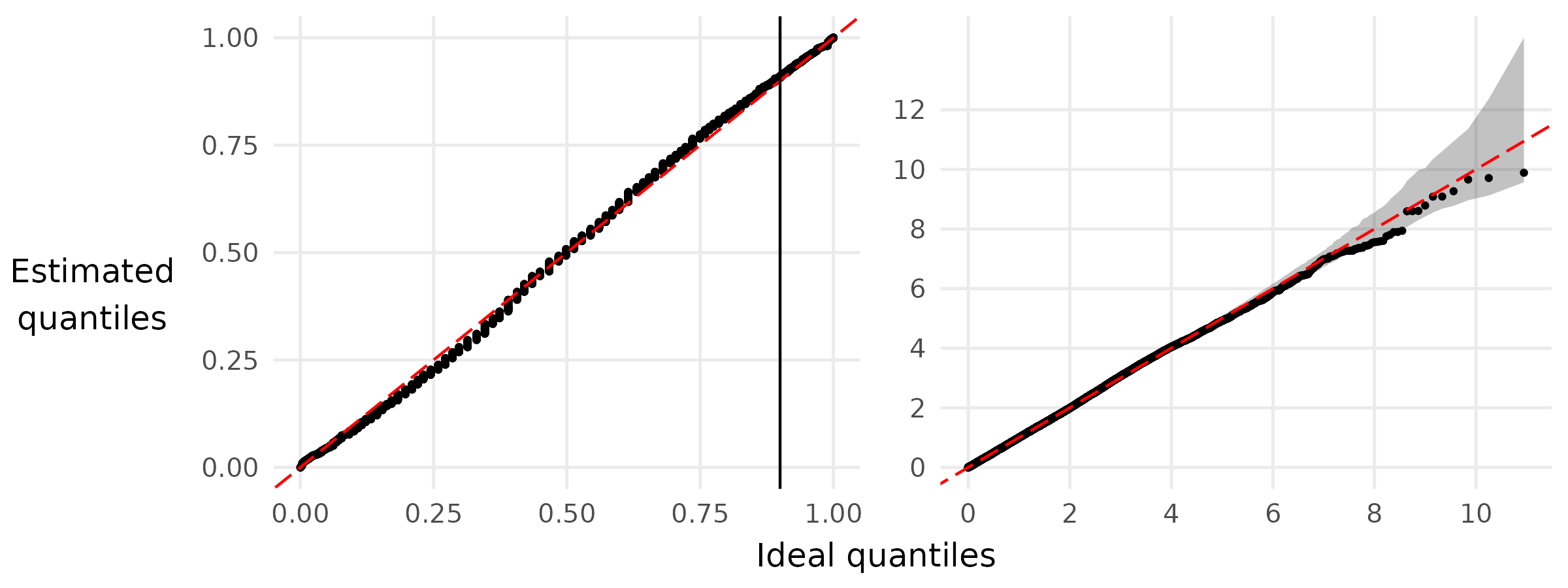}
 \caption{Spatially and temporally pooled QQ-plots for model $M_2$: (left) all data on uniform margins, threshold shown as vertical line; (right) tail model (GPD) on exponential margins. The shaded region shows pointwise 95\% tolerance intervals. The lines of equality are in red.}
 \label{fig:qqplots}
\end{figure}

The model selection diagnostics in Table~\ref{tab:gpd_models_and_cv_res} show primarily the relative quality of the three model fits. To assess the absolute quality of the fitted model $M_2$ we use pooled QQ plots in Figure \ref{fig:qqplots}, pooling over all sites and years. Due to the spatio-temporal non-stationarity of the marginal model, we transform the data through our fitted model into a common uniform scale and to a common exponential scale (for the conditional distribution of threshold excesses). The choice of scales helps identify key departures of fit in the body and tails of the distribution respectively. We see evidence of an exceptionally strong fit in both components of the distribution, with values near the lines of equality, and in the far upper tail, all values falling within the pointwise tolerance bounds which were derived assuming independence of time and space (so are much narrower than necessary). Examples of similar site-specific QQ plots are shown in \citet[Figure 12]{Healy2023sup} for the five stations identified in Figure~\ref{fig:obs_map} and for five other randomly selected stations. These show a slightly more varied quality of fit, with the least good fits occurring on the coastlines, e.g., Malin Head, but with very good fits at most stations.
\\
\section{Spatial models}\label{SEC4}
\subsection{Standardising data}\label{SEC4_1}
\noindent When modelling dependence between variables with differing marginal distributions and covariates, it is common to first standardise the marginal variables so that they have an identical distribution over variables and covariates \citep{Coles2001,DeHaan2006}. Here we transform the data to (unit) Pareto distributions, $X^P$, $X_c^P$ and $X^P_o$, using the same subscript notation as in Section~\ref{SEC3}. The choice of Pareto marginal scale is ideal for studying asymptotically dependent variables, a property defined in Section~\ref{SEC4_2},
such as $r$-Pareto processes \citep{DeHaan2006}, but less ideal for asymptotically independent variables, where shorter-tailed Exponential or Laplace distributions are favoured \citep{Wadsworth2019}. We use of the probability integral transform, i.e.,
\begin{equation}
\label{paretoMargins}
 X^P(t, \boldsymbol{s}) = 1/\left[1-F_{t, \boldsymbol{s}}\{X(t, \boldsymbol{s})\}\right], \mbox{ for all }
 \boldsymbol{s}\in \mathcal{S} \mbox{ and all }t
\end{equation}
where the marginal distribution function $F_{t, \boldsymbol{s}}$ takes a different estimated form below and above $u(\boldsymbol{s})$, see Sections~\ref{sec:below} and \ref{sec:above}. Thus, if the marginal model is perfectly estimated, we have $\Pr(X^P(t, \boldsymbol{s})>y)=y^{-1}$ for all $y>1$, $t$ and $\boldsymbol{s}$. In our uncertainty assessment in the subsequent inference, the marginal model uncertainty is accounted for through our bootstrap procedures.
To transform from standard Pareto margins to the original distribution at $\boldsymbol{s}$ and $t$, the inverse of the transformation~\eqref{paretoMargins} is used.\\

\subsection{Classification of extremal dependence type}\label{SEC4_2}
\noindent We now explore the nature of the extremal spatial dependence structure in the processes $X_c^P$ and $X^P_o$. To simplify notation we omit the temporal dimension of these spatial processes but always consider the process on the same day at different locations. Following 
\citet{Coles1999}, we estimate the pairwise coefficient of asymptotic dependence, $\chi$. Specifically, for the process $X^P$ at sites $\boldsymbol{s}_i$ and $\boldsymbol{s}_j$, $\chi= \chi^P(\boldsymbol{s}_i, \boldsymbol{s}_j)$
is defined by
\begin{equation}
\label{eq:chi_lim}
 \chi^P(\boldsymbol{s}_i, \boldsymbol{s}_j) = \lim_{v \to \infty} \Pr(X^P(\boldsymbol{s}_j) > v \big| X^P(\boldsymbol{s}_i)>v).
\end{equation}
If $\chi^P(\boldsymbol{s}_i, \boldsymbol{s}_j)>0$ (or equals 0) then
$X^P$ is said to be asymptotically dependent (or asymptotically independent) respectively for these sites. The larger the value of $\chi$ ($0 < \chi\le 1$) the stronger the asymptotic dependence.

The selection of the appropriate extremal dependence model for the data depends on whether or not the process is better approximated as being asymptotically dependent for all $\boldsymbol{s}_i,\boldsymbol{s}_j \in \mathcal{S}$ or not. The base quantity that is typically used to identify asymptotic dependence for a pair of sites is $\chi^P(\cdot, \cdot;p)$, where 
\begin{equation}
\label{eq:chi_subasy}
 \chi^P(\boldsymbol{s}_i, \boldsymbol{s}_j;p) =\Pr(X^P(\boldsymbol{s}_j) > v_p \big| X^P(\boldsymbol{s}_i)>v_p),
\end{equation}
with $v_p=1/(1-p)$ being the $p$th marginal quantile of $X^P$. An empirical estimate of $\chi^P(\boldsymbol{s}_i, \boldsymbol{s}_j;p)$ exploits the replication over $t$ by assuming spatial dependence does not change with $t$. We denote this estimator by $\tilde{\chi}^P(\boldsymbol{s}_i, \boldsymbol{s}_j;p)$. We expect approximate spatial stationarity and isotropy of the spatial extreme process. Plotting (not show) the cloud of $\tilde{\chi}^P(\boldsymbol{s}_i, \boldsymbol{s}_j;p)$ against the Euclidean distance between the sites ($h_{ij}=\left\lVert\boldsymbol{s}_i-\boldsymbol{s}_j\right\rVert$), reveals a decay with distance that is somewhat hidden by the sampling variation of the points, with the variation depending on the overlap in time of samples at the pairs of sites. A better empirical estimate of $\chi^P(h;p)$, the pairwise extremal dependence at separation distance $h$, exploits the property that it changes smoothly over $h$ and that we can obtain the sampling distribution of $\tilde{\chi}^P(\boldsymbol{s}_i, \boldsymbol{s}_j;p)$ through the bootstrap. Together these enable us to construct a weighted estimate $\tilde{\chi}^P(h;p)$ from the cloud of points (using pairs with $h_{ij}$ close to $h$) and obtain its sampling distribution. We used 500 bootstraps and 30 binned distances, each with an equal number of pairs of sites.

\begin{figure}[ht]
	\centering
	\includegraphics[width =12cm]{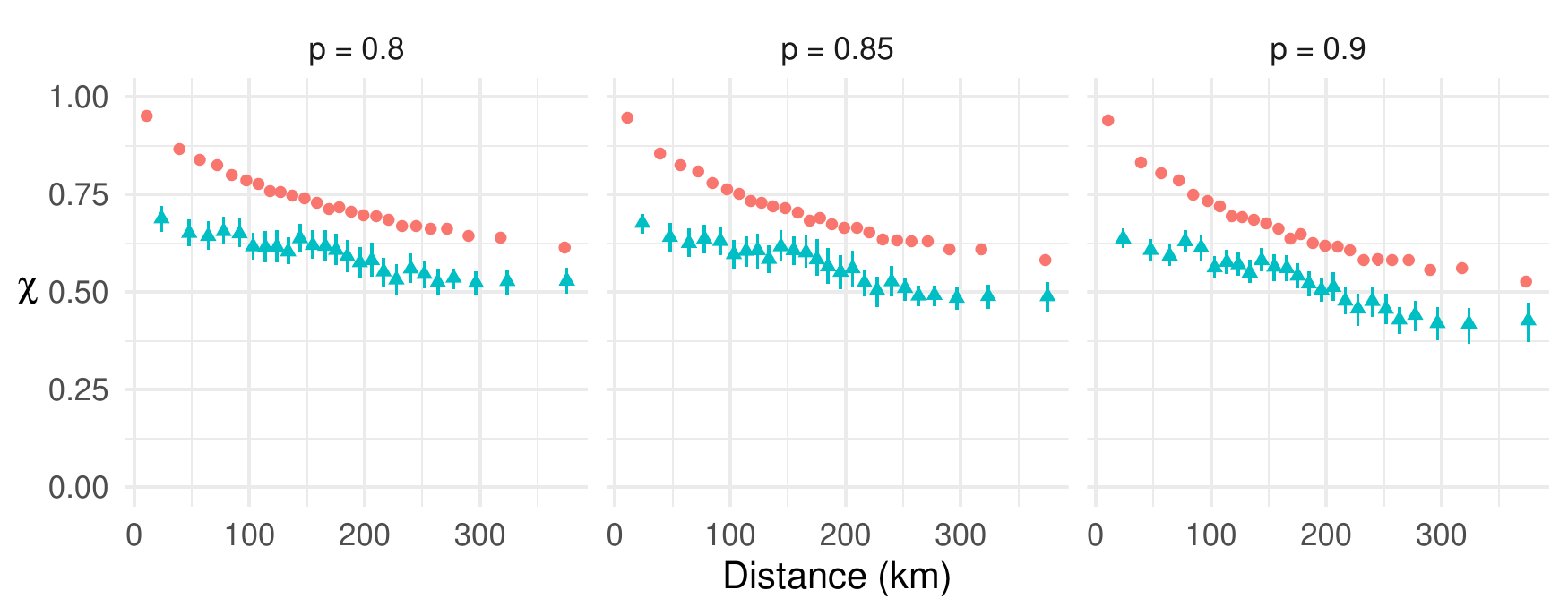}
	\caption{Empirical estimates $\tilde{\chi}^P(h;p)$ plotted against inter-site distance $h$ 
	for the climate model 
	$(\chi_c^{P})$ (orange) and station data $(\chi_o^{P})$ (blue). Plots use marginal model $M_2$ for $p=0.8, 0.85$ and $0.9$ (left to right) with 95\% confidence intervals shown as vertical lines. }
	\label{fig:chi_range}
\end{figure}

Figure~\ref{fig:chi_range} shows the behaviour of $\tilde{\chi}^P(h;p)$, for both
$X_c^P$ and $X^P_o$ processes. It shows estimates and intervals that account for 95\% 
of the marginal estimation uncertainty, which for $X_o$ we used model $M_2$. These
estimates are shown 
for $p \in (0.8, 0.85, 0.9)$, the latter corresponding to only 9 days per summer.
Despite the climate model having a much richer set of pairs of sites and longer simultaneous data, both processes provide very similar qualitative findings. Naturally, $\chi^{P}(h;p)$ decreases with distance but in all cases, it is far from zero, even at the longest distances of pairs of sites from $\mathcal{S}$. For short distance $\chi^{P}_c(h;p)$ there is more variation in the estimated values than for $\chi^{P}_o(h;p)$. We have that $\hat{\chi}^P_o(h;p)\le \hat{\chi}^P_c(h;p)$ for all distances, suggesting that the $X_c$ data are overestimating the extremal dependence in $X_o$. This difference is important
when looking at extreme events spatially, as it suggests using the climate model data alone (or when down-scaled) will lead to an overestimate of the risk of widespread heatwaves in Ireland.

Most critical for our modelling of the observed process is to assess whether, as $p$ increases to $1$, the $\hat{\chi}^P_o(h;p)$ values decay to zero or stabilises at a non-zero limit indicating asymptotic independence and asymptotic dependence respectively. There is a small decline, at all distances, however, even when $p=0.9$ these estimates are far from zero for both $X^P_c$ and $X^P_o$. So we conclude that it seems reasonable that maximum daily temperature data are consistent with asymptotic dependence over Ireland.\\

\subsection{$r$-Pareto Processes}\label{SEC4_3}
\noindent We now model the extreme values of the process $X^P(t,\boldsymbol{s})$ over $\boldsymbol{s} \in \mathcal{S}$, with unit Pareto distributed marginal variables. We look at spatial fields separately for each $t$ and simplify notation by dropping the argument $t$. First, we define what we mean by a spatially extreme event as are there is no natural ordering of multivariate or spatial processes. Here the level of extremity of the stochastic process $X^P:=\{X^P(\boldsymbol{s}): \boldsymbol{s}\in \mathcal{S}\}$
is determined by a risk function $r(X^P)\in [0,\infty)$, where the only constraint on $r$ is that it is homogeneous of order 1, i.e., $r(c \boldsymbol{x})=c r(\boldsymbol{x})$ for any constant $c>0$ and with $\min(\boldsymbol{x})>0$. \citet{DeFondeville2018a, DeFondeville2022} suggest taking $r$ as the magnitude at one particular site of interest, or the spatial mean, median, maximum or the minimum over $\mathcal{S}$.

Under weak conditions on $X^P$, \cite{DeFondeville2018a} report that 
\begin{equation}\label{rparetoeq}
 \operatorname{Pr}\left\{ \left[v^{-1}X^P(\boldsymbol{s}) : \boldsymbol{s} \in \mathcal{S} \right] \in \cdot \mid r(X^P)>v\right\} \rightarrow \operatorname{Pr}\left\{ \left[Y^P_r(\boldsymbol{s}) : \boldsymbol{s} \in \mathcal{S} \right] \in \cdot\right\}
\end{equation}
as $v\rightarrow \infty$, where $\left\{Y^P_r(\boldsymbol{s}): \boldsymbol{s} \in \mathcal{S}\right\}:= Y^P_r$ is marginally non-degenerate for all $\boldsymbol{s} \in \mathcal{S}$, with $Y^P_r$ a $r$-Pareto process. Limit~\eqref{rparetoeq} implies that scaled events of the $X^P$ process with risk exceeding a threshold of $v$ are increasing well-approximated by an $r$-Pareto process, as the risk threshold increases to infinity. 
The limit~\eqref{rparetoeq} is used for statistical modelling by taking it as an equality for 
a suitably large value for $v$, denoted by $v_r$ with $v_r>1$, then those spatial events with a risk function exceeding $v_r$ are treated as realisations from an $r$-Pareto process. Specifically, taking an extreme set $B\subseteq \{\boldsymbol{x}: r(\boldsymbol{x})>1\}$ 
leads to the modelling assumption that
\begin{equation}\label{rparetoeq2}
 \operatorname{Pr}\left\{ X^P \in v_rB \mid r(X^P)>v_r\right\} = \operatorname{Pr}\left\{Y_r\in B\right\}. 
\end{equation}
Hence defining the set $A=v_rB$ and un-doing the conditioning on the left-hand side of equality~\eqref{rparetoeq2}, 
for any $A\subseteq \mathcal{A}_r:=\{\boldsymbol{x}: r(\boldsymbol{x})>v_r\}$ we that
\begin{equation}\label{rparetoeq4}
 \operatorname{Pr}\left\{ X^P \in A \right\} = \Pr\{r(X_r^P)>v_r\}\operatorname{Pr}\left\{v_rY_r\in A\right\}.
\end{equation} 

The $r$-Pareto process exhibits properties 
which can be exploited for efficient evaluation of $\operatorname{Pr}\left\{ X^P \in A \right\}$.
Specifically, $Y^P_r$ decomposes into two independent components:
\begin{equation}\label{decomp}
 Y^P_r(\boldsymbol{s}) = RW_r(\boldsymbol{s})
 \mbox{ for all }\boldsymbol{s} \in \mathcal{S},
\end{equation}
where $R$ is unit Pareto distributed and is interpreted as the risk of the process, and $W_r:=\{W_r(\boldsymbol{s}): \boldsymbol{s} \in \mathcal{S}\}$ is a stochastic process which describes the spatial profile of the extreme event, i.e., the proportion of the risk function $r$ contributed by each site. The limiting dependence structure of $X^P$ is entirely determined by the stochastic properties of $W_r$. By construction $R = r(Y^P_r)$ and $r(W_r) = 1$. A consequence of the limiting 
approximation~\eqref{rparetoeq2} holding above $v_r$ and $R$ having a Pareto distribution is that expression~\eqref{rparetoeq4} simplifies as we have $\Pr\{r(X^P)>v_r\}=v_r^{-1}$.

The characterisation~\eqref{decomp} is powerful for the extrapolation to larger events than those observed 
due to $R$ having a known distribution and the independence property ensuring that the spatial profiles of larger events have exactly the same stochastic properties for any event with a risk greater than 1. For any $r$-Pareto process and a set $A\subseteq \mathcal{A}_r$, 
there exists a constant $b_A\in [1, \infty)$ such that any $b\in [1,b_A]$ we have that
 \begin{equation}
 \label{eqn:scaling}
\operatorname{Pr}\left\{v_rY_r \in A \right\}=b^{-1} \operatorname{Pr}\left\{bv_rY_r \in A \right\}. 
\end{equation}
 Although the two sides of this expression are equal, the two probabilities are not, with \citet{Opitz2021} noting that the latter is much more efficient to estimate using Monte Carlo methods. Taking $b>b_A$ will give bias, as some smaller outcomes in $A$ will be missed by simulations of $bv_rY_r$ 
while $b<b_A$ leads to unnecessary variability in the empirical estimator. So we look to scale by $b_A$ in Section~\ref{sec:SimProc}, where we discuss how to obtain $b_A$ and illustrate its usage in estimating the right-hand side of expression~\eqref{rparetoeq4}.

The above shows that inference for any extreme events is relatively straightforward once we have a model for the process $W_r$. We follow \citet{Engelke2015} and \citet{DeFondeville2018a}
by modelling $W_r$ as a spatial stationary isotropic log-Gaussian stochastic process which is determined solely by a variogram $\gamma(h)$, for inter-site distance $h\ge 0$. We use the Matérn variogram family 
\begin{equation}
\label{eqn:mat}
 \gamma_{\text{mat}}(h;t) = 
 \alpha \left\{1 - (2 \sqrt{\nu} h /\phi)^\nu
 2^{1-\nu}\Gamma(\nu)^{-1} \text{K}_\nu (2 \sqrt{\nu} h/\phi) \right\},
\end{equation}
where $K_\nu$ is a modified Bessel function of the second kind and the positive parameters $(\alpha=\alpha_t, \phi=\phi_t,\nu=\nu_t)$ determine the variance, range and smoothness respectively at time $t$. Our choice of a bounded variogram was based on the evidence from Figure~\ref{fig:chi_range} which suggested that the summer temperature process is asymptotically dependent, even at the longest distances in Ireland, see \citet[Section 7]{Healy2023sup} for additional discussion on the choice of variogram function including the support for isotropy. 

The issue of missing data in spatial extremes applications 
seems to be rarely discussed. A possible reason for this is that with composite likelihood fitting methods for max-stable processes \citep{Padoan2010} the implications are restricted as only pairwise joint likelihood contributions are used so the impact of missing data is limited. This is not the case for $r$-Pareto processes, which model jointly across all sites, and the issue of missing data in this context does not seem to have been discussed. When encountering missing data it is tempting to remove all observations at that time point from all sites in the network. However, we observed in Section~\ref{SEC2} that this tactic would leave us with no data! Fortunately, thanks to the properties of the log-Gaussian process, it is possible to show that the model is closed under marginalisation \citep{Engelke2020}.

With missing data, we need to be careful in selecting a suitable risk function $r$. The choice of risk function needs to be invariant to the changing dimension of partially observed events, whatever their missing patterns. Hence in our statistical inference at time $t$, for all $t\in \mathcal{T}_o$, we take the risk function $r_t$ 
to be the average of the standardised variables over the stations which were observed at time $t$, i.e., 
\begin{equation}
r_t(X_o^P(t,\boldsymbol{s}): \boldsymbol{s}\in \mathcal{S})
=\sum X_o^P(t,\boldsymbol{s}_i)I_o(t, \boldsymbol{s}_i)\big/ \sum I_o(t, \boldsymbol{s}_i), 
\end{equation}
where $I_o(t, \boldsymbol{s}_i)$ is the indicator variable for whether $X_o(t,\boldsymbol{s}_i)$ is observed or not, and the sums are from $i = 1, \dots, |\mathcal{S}_o|$. For evaluating $r_t$ we would have liked to use a subset of stations that are observed for all $t$ and reasonably evenly spread across Ireland. This was not possible, however, the data from Aldergrove (north) are used as this site has very little missing data over the period 1930-2020.

We set the risk threshold $v_r$, used inequality~\eqref{rparetoeq2} to define extreme spatial events, at the 80\% sample quantile of the risk values calculated from all observed events, i.e., we use the empirical estimate $\widetilde{\Pr}\{r(X_r^P)>v_r\}=0.2$.
We explored different threshold choices and selected the lowest level we could whilst making the usual bias/variance trade-off for tail selection. \citet[Figures 15 and 16]{Healy2023sup} show that the parametric estimate of $\chi^P_o$, derived from the variogram, agrees well with empirical estimates for each marginal model $M_0, M_1$ and $M_2$.

We explored the effect of a time-changing dependence model. We allowed for the variance and range parameters of the Matérn variogram~\eqref{eqn:mat} to vary over time $t$, while keeping $\nu$ constant. We considered a range of constant and log-linear models using each of the marginal models $M_0$ to $M_2$. For $M_0$ we find some evidence, at the $5\%$ level, for $\alpha_t$ increasing with $M^I(t)$ (weakening dependence over time), but not for the improved marginal models $M_1$ or $M_2$. Evidence for a change in extremal dependence was not statistically significant and so we keep a temporally stationary $r$-Pareto process. \\

\subsection{Simulation and efficient inference for spatial extreme events}
\label{sec:SimProc}
\noindent We simulate spatial extreme events on the observational scale in year $t$ by first simulating an event from the $r$-Pareto process and then map this pointwise to the data scale using the inverse of transform~\eqref{paretoMargins} for the required $t$. For each step of this process we use the selected statistical model and the simulated values are generated using the fitted parameters of that model, or for assessing uncertainty in the point estimates, using the bootstrapped realisations of these parameters. As the 
estimated $r$-Pareto process in our application is found to be well-approximated by a stationary process over time, we can generate identically distributed events of the $r$-Pareto process to transform for each location and time $t$ using the time-varying marginal model. The $r$-Pareto process simulations are generated using the R package \texttt{mvPot} \citep{DeFondeville2021}. We denote these simulations by $\boldsymbol{y}^P_1,\boldsymbol{y}^P_2, \dots, \boldsymbol{y}^P_m$, for $m$ simulations, with the
$i$th simulation consisting of the spatial realisation
$\boldsymbol{y}^P_i=\{y^P_i(\boldsymbol{s}): 
~\boldsymbol{s}\in \mathcal{S}\}$. For the $i$th realisation of the $r$-Pareto process, $\boldsymbol{y}^P_i$, we define , $r_i=r(\boldsymbol{y}^P_i)>1$ as the risk, 
$\boldsymbol{w}_i=\boldsymbol{y}^P_i/r_i$ as 
the spatial profile, and 
$w_i(\boldsymbol{s})=y^P_i(\boldsymbol{s})/r_i$ 
as the value of the $r$-Pareto process at site $\boldsymbol{s}\in \mathcal{S}$.

\citet[Figure 20]{Healy2023sup} shows five simulated extreme events, transformed to observational scale under 2020 conditions and the exact same events in 1942 conditions (presented as a difference in temperatures at each site, for each event). A positive difference shows the equivalent event in the two years to be hotter in 2020 than 1942, with that difference found to be largest for the hotter events.
As the $r$-Pareto process realisations can have marginal values not in the range $(0,1)$ at some sites, i.e., outside the domain of the Pareto variable we follow \citet{DeFondeville2018a} and for transformation to the observational space we use Fr\'echet marginals, not Pareto.

Although expression~\eqref{rparetoeq4} provides a basis for inference for the probability of occurrence in any extreme event $A\subseteq \mathcal{A}_r$ by the process $X^P$, we are most interested in making inference about spatial events of the observational process that exceed a critical temperature of $T^\circ$C somewhere over Ireland at time $t$. We denote these events by \begin{align}\label{eqn:ExtremeSet1}
 A_{t,\mathcal{S}}(T)&=\{X_o(t, \boldsymbol{s}), \boldsymbol{s} \in \mathcal{S}: \exists ~\boldsymbol{s}_0\in \mathcal{S} \mbox{ with }X_o(t, \boldsymbol{s}_0)>T\}.
\end{align}
After the marginal transformation to Pareto margins, this event is equal to 
\begin{align}\label{eqn:ExtremeSet2}
 A_{t,\mathcal{S}}(T)&=\{X^P_o(t,\boldsymbol{s}), \boldsymbol{s}\in \mathcal{S}: \exists ~\boldsymbol{s}_0\in \mathcal{S} \mbox{ with }X^P_o(t,\boldsymbol{s}_0)>T^P(t, \boldsymbol{s}_0)\}
\end{align}
where $T^{P}(t, \boldsymbol{s})$ is the mapping of $T$ through the transformation~\eqref{paretoMargins}
at time $t$ and for site $\boldsymbol{s}$. To use the r-Pareto approximation all elements of $A_{t,\mathcal{S}}(T)$ must have a risk exceeding $v_r$; this imposes a lower bound $T>20.6^\circ$C across all $t$. We also focus on marginal extreme events, which restricts $T\ge \max_{i\in \mathcal{S}}u_i=22.9^{\circ}$C. As all the results we present in Section~\ref{SEC5} are for $T\ge 26^{\circ}C$ this lower bound is not restrictive for our purposes. 

To estimate $\Pr\left\{A_{t,\mathcal{S}}(T) \right\}$ there have been a set of possibilities proposed, see \citet[Section 6.1]{Healy2023sup}. We focus on the most efficient of these estimators, which exploits the independence property~\eqref{decomp}, and the scaling property~\eqref{eqn:scaling}. 
Specifically, $\boldsymbol{w}_i$ and $r_j$ are independent realisations for all $i,j$, and there is no reason to restrict ourselves to the observed $r_j$ as we know they are unit Pareto realisations. So we supplement the information to have $\{r^P_{j}; j=1, \ldots ,L\}$, which are i.i.d.~realisations of a unit Pareto variable, where $L$ is taken as large as possible to improve computational efficiency. To find the optimal scaling factor $b_{T(t)}$ we first define component-wise maxima of the simulated Pareto processes scaled to have unit cost, i.e., $\omega_{(m)}(\boldsymbol{s}) = \max_{i = 1, \dots, m} w_i(\boldsymbol{s})$, for each $\boldsymbol{s}\in \mathcal{S}$. At time $t$, we want to scale these component-wise maxima by as much as possible without producing a scaled event with an exceedance of $T^P(t,\boldsymbol{s})$ for some $\boldsymbol{s}\in \mathcal{S}$. The appropriate scaling is then $b_{T(t)} = \min_{\boldsymbol{s}\in \mathcal{S}} \left\{ T^P(t, \boldsymbol{s}) / \omega_{(m)}(\boldsymbol{s}) \right\}$. 
Here $b_{T(t)}=v_rb_A$ in expression~\eqref{eqn:scaling}. Combined together these give a form of importance sampling estimator
\begin{equation} 
\hat{\Pr}_{imp}\left\{ A_{t,\mathcal{S}}(T)\right\}
 = 
 \frac{1}{b_{T(t)}mL}\sum_{i = 1}^m
 \sum_{j = 1}^L
 I\left\{\exists \boldsymbol{s_0} \in \mathcal{S}: r^P_j b_{T(t)}
\frac{y^P_i(\boldsymbol{s_0})}{r_i} > T^P(t, \boldsymbol{s_0}) \right\},
 \label{eqn:imp}
\end{equation}
see \citet[Algorithm 2]{Healy2023sup}. With this scaling choice, and the extrapolation from the $r^P_j>\max(r_1, \ldots r_m)$, we are guaranteed to have at least $m$ out of the $mL$ simulated fields which achieve at least temperature $T^{\circ}C$ somewhere in $\mathcal{S}$
in year $t$. \\

\section{Temporal changes in spatial extreme events}\label{SEC5}
\begin{figure}[ht]
 \centering
 \includegraphics[width = 15cm]{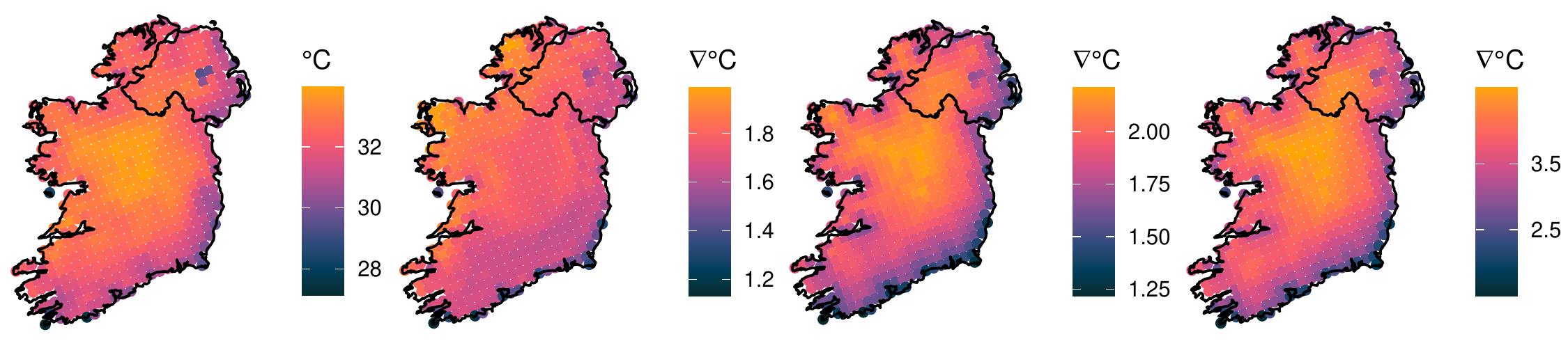}
 \caption{(1st plot) Estimated 100-year marginal return level for the year 2020; (3rd plot) estimated change in 100-year marginal return level from 1942 to 2020; (2nd and 4th plots) lower and upper 95\% CI limits respectively for the change in 100-year return level from 1942 to 2020.}
 \label{fig:100_yr_rl}
\end{figure}

\noindent We present a range of summaries detailing how spatial daily maximum temperature extreme events in Ireland are changing over the period 1942-2020.
First, we look at the changes in the marginal quantiles. Figure~\ref{fig:100_yr_rl} shows estimates of the level exceeded by daily values with probability $1/9200$, i.e., that of a 100-year return level if the process was stationary in time. For simplicity, we refer to these as the 100-year levels changing over time. For model $M_2$ we show these estimates for 2020 and also present the estimated difference between them for 2020 and 1942. In the latter, a positive value represents a warming of temperatures. The 100-year return level has increased between $1.2-2.2^\circ$C across Ireland, with the larger increases away from the south and east coasts. 
These changes in extreme temperatures over the observed record are substantially larger than the $1^\circ$C change of
$M^I$ and $M^G$ over this period illustrating that climate change is more radically affecting summer temperature extreme events than mean levels.

In Figure~\ref{fig:100_yr_rl} we also report
the change in the upper and lower 2.5\% quantile of change in return level from 1942 - 2020 across all bootstrap samples. In both cases and at all sites these changes are positive, with the rate of change in these features being greater than that of the point estimates. Although we present the results for the 100-year return levels, similar results hold for all high quantiles and for the finite upper endpoints of the marginal distribution; the latter as the GPD shape parameter being negative (see Table~\ref{tab:gpd_models_and_cv_res}). 
\citet[Figures 13 and 14]{Healy2023sup} gives equivalent figures for $M_0$ and $M_1$. 

\begin{figure}[ht]
 \centering
 \includegraphics[width = 7cm]{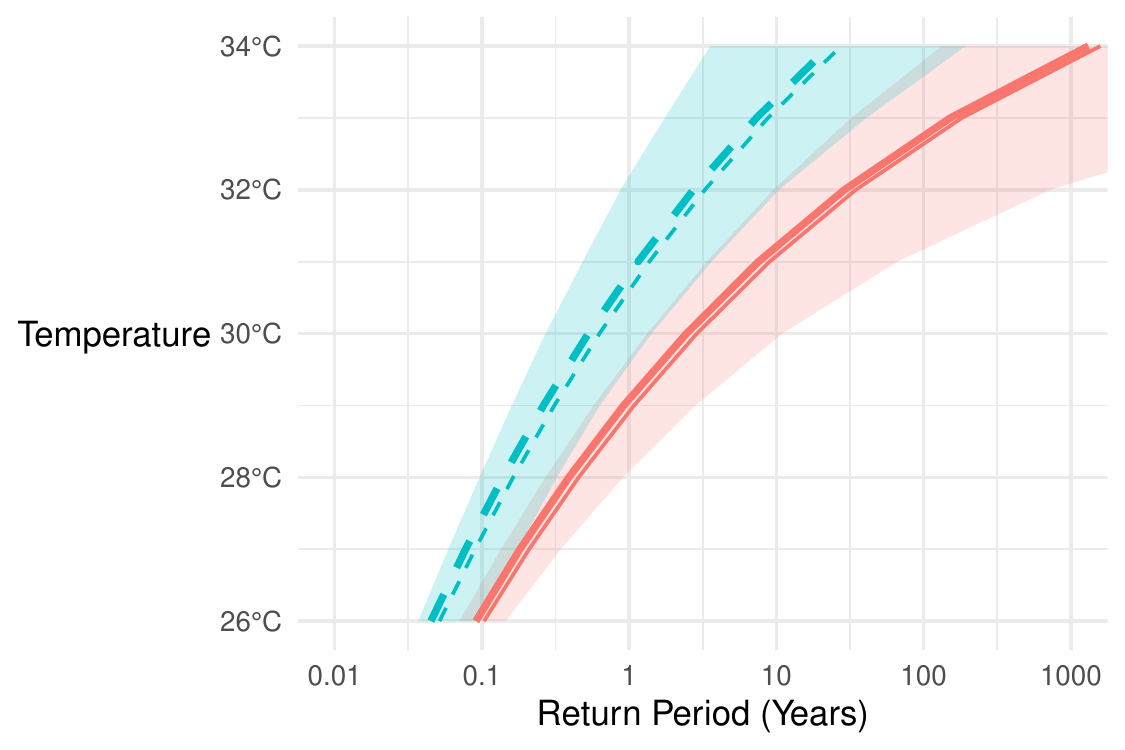}
 \caption{Return period of the event $A_{t,\mathcal{S}}(T)$ where an extreme temperature exceeding $T^{\circ}$C occurs somewhere on the Irish station network, $\mathcal{S}_o$. Blue dashed (solid orange) lines correspond to $t=2020$ ($1942$). Shaded regions show pointwise 95\% confidence intervals for the return periods. The higher bold curves show the corresponding point estimates for the climate model grid $\mathcal{S}_c$.}
 \label{fig:spatialRL}
\end{figure}

Next, we consider summaries that also reflect the dependence structure of extreme events. There are no established analytical closed-form expressions of such changes. Instead, we revert to using simulated fields of extreme events and presenting risk measures based on empirical summaries using large samples of these fields. The simulation strategy set out in Section~\ref{sec:SimProc}
gives replicated independent spatial fields. 
In particular, we focus on the occurrence of events $A_{t,\mathcal{S}}(T)$, i.e., an extreme temperature of at least $T^{\circ}$C somewhere in Ireland (determined by the set of locations as required), and then summaries of the properties of such events.
We estimate $\Pr\{A_{t,\mathcal{S}}(T)\}$ using the estimator $\hat{\Pr}_{imp}$, with $m = 25,000$ and $L = 300$. Figure~\ref{fig:spatialRL} shows this estimated probability (expressed as a return period) for a range of temperatures $T \in [26,34]$ for years 1942 and 2020 separately for $\mathcal{S}_o$ and $\mathcal{S}_c$. For $\mathcal{S}_o$,
the plot reveals a marked change with estimated return periods being shorter in 2020 compared to 1942 for the same $T$. To illustrate this, consider
the event with the hottest temperature observed anywhere at the station network, a temperature of $33^{\circ}$C at Phoenix Park, Dublin, July 2022. The spatial event $A_{t,\mathcal{S}}(33)$ changes from being a 1 in 182-year event in 1942 to a 1 in 8.7-year event in 2020. Furthermore, the model estimates that a temperature in excess of $34^{\circ}$C, i.e., a value not yet recorded in Ireland, changes from a 1 in 1,588-year event to a 1 in 27.5 year event over this time window.

Figure~\ref{fig:spatialRL} shows that, for a given return period, hotter temperatures are expected somewhere in $\mathcal{S}_c$ and on
$\mathcal{S}_o$, as the former is denser and with better coverage than the station network. The difference between the results for the two collections of sites is very small. This slight change shows that the station network, when all gauges are working, has the ability to fully capture all extreme temperature events over Ireland. 
Such information has not been available previously given the complexity of addressing spatial dependence, marginal non-stationary and missing data in the station network.

We now propose risk metrics to summarise the features of events satisfying $A_{t,\mathcal{S}}(T)$. First consider a measure of pairwise dependence, which extends the idea behind $\chi^P$ but applied to the data scale, so it 
combines the effects of changes over time in the marginal distributions and the estimated extremal dependence structure. Specifically, we define
\begin{equation*}
\label{eq:chi_limObs}
 \chi_o(h;A_{t,\mathcal{S}}(T)) 
 = \Pr\{X_o(t,\boldsymbol{s}^h) > T ~\big| ~ A_{t,\mathcal{S}}(T)\},
\end{equation*}
where $\boldsymbol{s}^h$ is a randomly selected site in $\mathcal{S}$ with $\left\lVert\boldsymbol{s}^h-\boldsymbol{s}_0\right\rVert=h$, i.e., the conditional probability of the observational process exceeding temperature $T$ on day $t$ at a site which is a distance $h$ away from a site $\boldsymbol{s}_0$ that has a temperature
 exceeding $T$ on that day. We also investigate the associated unconditional risk measure
\begin{equation*}
\chi_o(h;T, t)= \Pr(\exists ~\boldsymbol{s}_0 \in \mathcal{S}: \min[X_o(t,\boldsymbol{s}_0),X_o(t,\boldsymbol{s}^h)] > T). 
\end{equation*}

\begin{figure}[ht]
 \centering
 \includegraphics[width = 12cm]{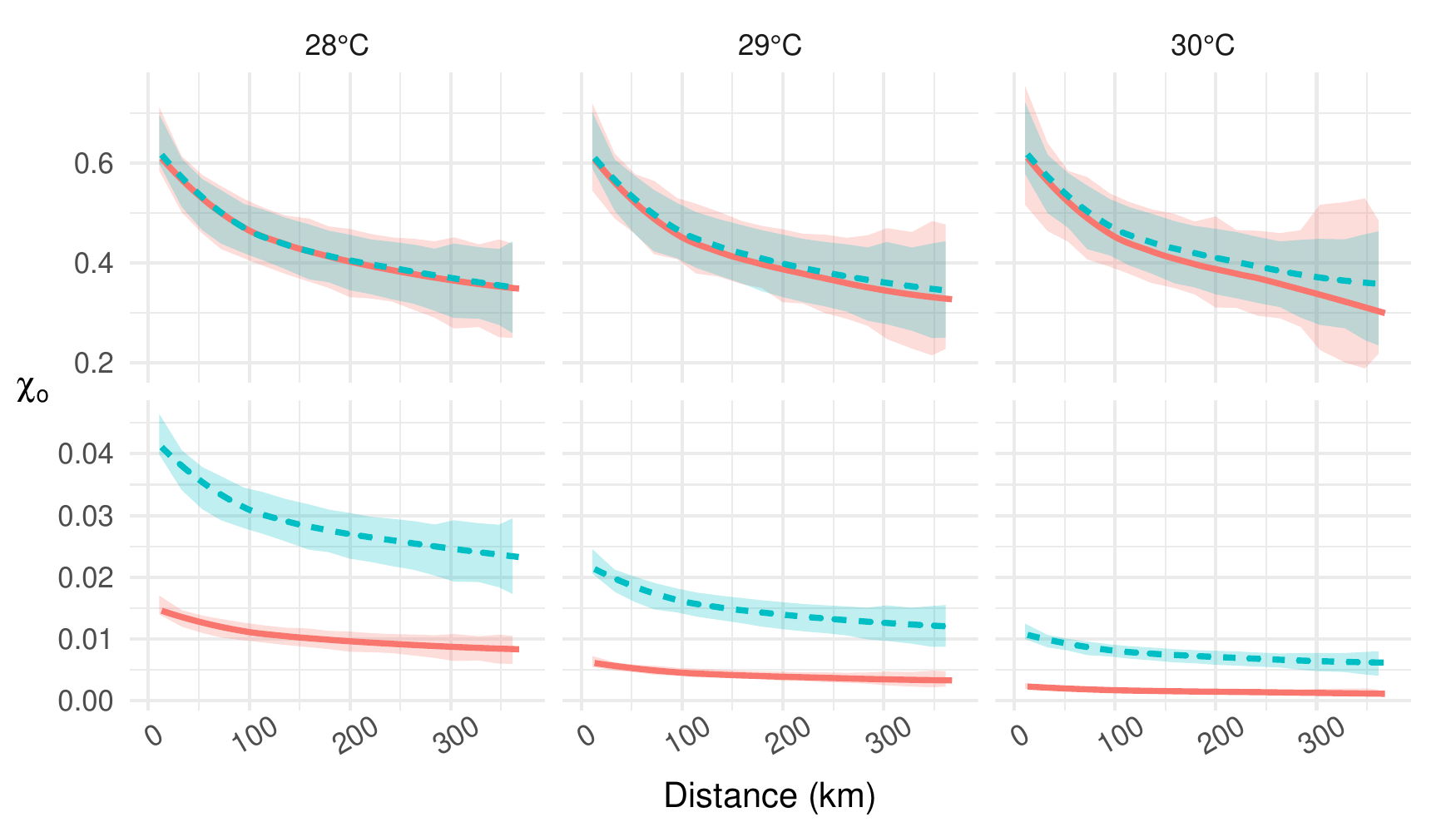}
 \caption{Estimates of 
 $\chi_o(h;A_{t,\mathcal{S}}(T))$ (top row) and $\chi_o(h;T, t)$ (bottom row) against $h$ (in km) for $T=28,29$ and $30^\circ$C for 1942 (solid, orange line) and for 2020 (dashed, blue line) for model $M_2$. Confidence intervals are based on 10,000 simulations for each 500 bootstrap sample datasets.}
 \label{fig:chi_orig_scale}
\end{figure}

Figure~\ref{fig:chi_orig_scale}
presents estimates of each of these two risk measures for a range of $h$ and $T$, between 1942 and 2020. On the observed data scale, we find that extremal spatial dependence decreases with distance as would be expected, but beyond this, there
are quite different findings from the two measures.
Risk measure $\chi_o(h;A_{t,\mathcal{S}}(T))$ 
is broadly stable over the presented range of $T$ and $t$ whilst the unconditional $\chi_o(h;T, t)$ is substantially different. The former is perhaps not too surprising given that the model is asymptotically dependent (the dependence structure is invariant to any change in the extremity of an event) and the extremal dependence structure is estimated to be stationary over time. However, as the event is on the marginal scale and the marginal distributions are changing with time this finding was not anticipated. For $\chi_o(h;T, t)$ we do see that the joint probability of temperature being above $T$ at sites $h$ apart changes notably with time, e.g., taking $h=100$ km, we find that $\chi_o(h;T, t)$ has increased by a factor of 2.8, 3.5 and 4.7 for $T=28,29$ and $30^\circ$C respectively.

Finally, we look at a spatial risk measure based on the proportion, $C$, of a spatial field over the network that exceeds $T^\circ$C at time $t$. Specifically, we consider the expected value of $C$, denoted $E_{o}(C; t,T)$. We also consider the conditional expected value of $C$, given by $E_o\{C \mid A_{t,\mathcal{S}}(T)\}$, i.e., given that we have observed a temperature somewhere at the station network. This conditional expectation is closely related to a functional used in characterising heatwave events \citep{Cebrian2022}. Specifically, these functionals and their relationships are given as follows:
\begin{eqnarray*}
\label{eq:chi_expected_Obs}
 E_{o}(C; t,T) &=& \mbox{E}\left( \frac{1}{|\mathcal{S}|}\int_{\mathcal{S}}I\{X_o(t,\boldsymbol{s}) > T\}d\boldsymbol{s} \right)\\
&=& \mbox{E}\left( \frac{1}{|\mathcal{S}|}\int_{\mathcal{S}}I\{X_o(t,\boldsymbol{s}) > T\}d\boldsymbol{s} ~\Big\lvert ~A_{t,\mathcal{S}}(T)\right)\times \Pr
\{A_{t,\mathcal{S}}(T)\}\\ 
&=& E_o\{C \mid A_{t,\mathcal{S}}(T)\}
\times \Pr\{A_{t,\mathcal{S}}(T)\}, 
\end{eqnarray*}
where $I(B)$ is the indicator function of event $B$. 

\begin{figure}[ht]
 \centering
 \includegraphics[width = 10cm]{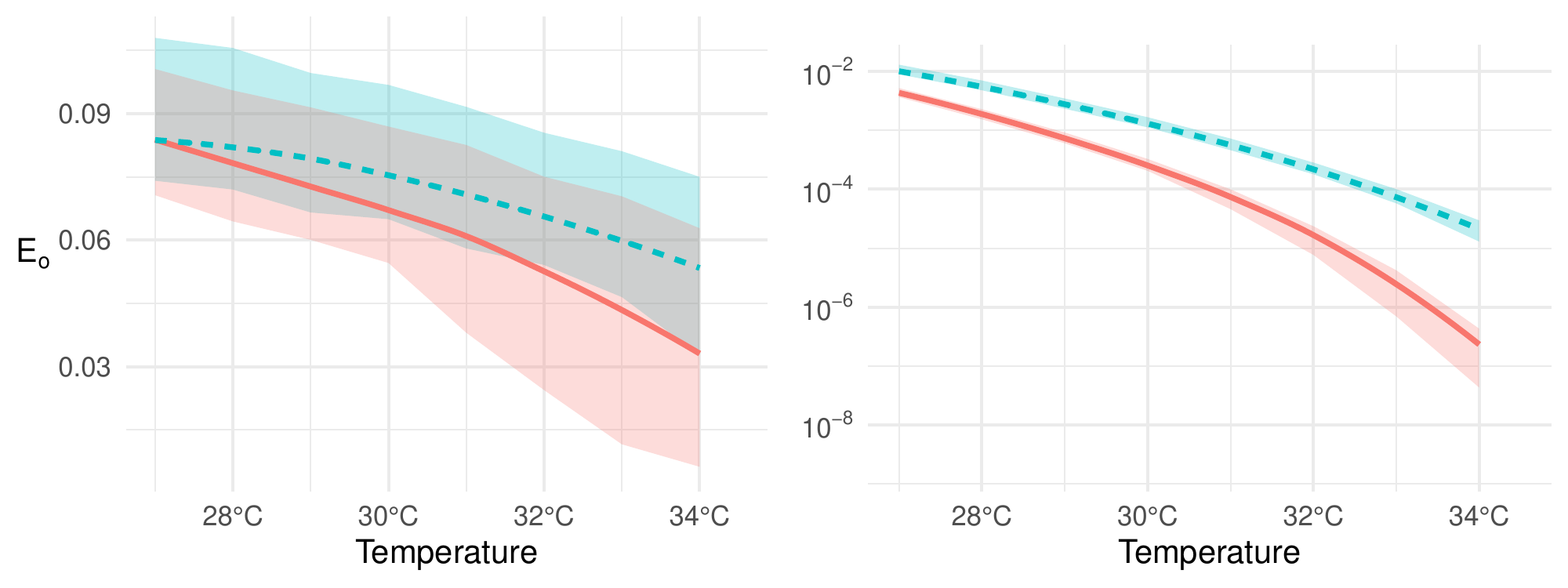}
 \caption{Left: expected proportion, $E_o\{C \mid A_{t,\mathcal{S}}(T)\}$, of Ireland that exceeds a temperature of $T^\circ$C in an extreme event given that at least one site in Ireland (at the station network) exceeds $T^\circ$C according to $M_2$. Right: the equivalent unconditioned estimates, i.e., estimates of $E_{o}(C; t,T)$. Estimates are plotted against $T$ for 1942 (solid, orange line) and for 2020 (dashed, blue line). The shaded regions give associated pointwise $95\%$ confidence intervals, based on $10,000$ simulated fields for each $500$ bootstrap sample dataset.}
 \label{fig:my_labelS}
\end{figure}

Figure~\ref{fig:my_labelS} shows that estimates of both of these measures for the station network over Ireland have increased from 1942 to 2020. The changes are highly significant, a factor of 90 larger for $T=34^{\circ}C$ when considering the unconditional expectation $E_{o}(C; t,T)$. However, when conditioning on the event $A_{t,\mathcal{S}}(T)$ this expected coverage proportion exhibits more limited changes with the largest difference being a doubling of the expected area affected when $T=34^{\circ}C$. In this latter case, the estimated change is small by comparison with its associated uncertainties.

We finish by reporting on our investigation of the sensitivity of our risk measure analyses to our marginal modelling choices for temporal non-stationarity. The results for the models $M_0$ and $M_1$, i.e., the less well-fitting models, are given in \citet[Section 6.2]{Healy2023sup}, with these 
being given for the same features shown for model $M_2$ in Section~\ref{SEC5}. Unsurprisingly, the inclusion of temporal non-stationarity in the tail model gives markedly different conclusions for all risk measures compared to those derived from the stationary model $M_0$. The inclusion of a coastal proximity covariate in model $M_2$ leads to larger scale parameter estimates inland and lower estimates in coastal regions than $M_1$, see \citet[Figure 7]{Healy2023sup}. This is reflected in the estimates of $\Pr\{A_{t, \mathcal{S}}(T)\}$ and $E_{o}(C; t,T)$, with $M_0$ being lower than both $M_1$ and $M_2$, and in $M_1$ giving slightly higher estimates than $M_2$ on the station network. This change results in the probability of observing $33^\circ$C somewhere in $\mathcal{S}_o$ increasing by a factor of 1.5, 25 and 21 between 1942 and 2020 for models $M_0, M_1$ and $M_2$ respectively, showing that the difference in key conclusions is not too large between models $M_1$ and $M_2$.\\

\section{Conclusions}\label{SEC6}
\noindent We have presented some novel candidate approaches to merge information from spatially and temporally complete climate models into the spatial extreme value analysis of sparse and temporally incomplete observed temperatures from available meteorological stations. New methodological features include using outputs from an extreme value analysis of the climate model data to provide a covariate for the equivalent analysis of observational data, and dealing with $r$-Pareto processes in a missing data framework. We also presented novel metrics, combining both marginal and dependence features, to describe changes in spatial risk over time.

Our analysis was for daily maximum summer temperatures over Ireland. We found that the climate model data were more helpful for marginal modelling of the observational data than for dependence modelling, as they have the potential to overestimate extremal spatial dependence relative to the observational data. We pooled data from across stations to fit our model and found evidence that the 
Irish summer temperature anomalies were the best-fitting covariate, appearing to mostly affect marginal behaviour with minimal effect on spatial extremal dependence, see \citet[Secion 2 and 7.2]{Healy2023sup}. We found that from 1942 to 2020 the occurrence rates of high threshold exceedances have increased by 35\%, with 95\% confidence interval 28-44\%, and extreme quantiles have increased by $1.2 - 2.2^\circ$C, the latter $\sim 1^\circ$C greater than the change of mean summer temperature anomalies for Ireland and globally. Finally, we found that spatial heatwave events over thresholds that are critical for society have become much larger, having at least doubled in spatial extent for 28°C, with this change increasing at more extreme temperatures.\\

\section{Discussion}\label{SEC7}

\subsection{Use of climate models}
\noindent In the analysis of climate extremes, practitioners tend to make use of some combination of observations and climate models, but without necessarily recognising the inherent trade-offs between them and the respective limitations in the data sources. Perhaps more critically we believe further thought could be given to the synergies which might enable a better set of tools for practitioners. Downscaling from climate models to match observational properties is a major area in statistical climate science, but it risks introducing biases in spatial dependence from climate models given its focus on the marginal agreement. Here we have illustrated what appears to be an effective new approach aiming instead to enhance the observational analysis by exploiting the strong information about the physical properties of the climate system and the greater spatial coverage of information embedded in the climate model data. 

As a proof of concept, we have restricted the information we extract from the climate models to that arising from one climate model output. There exists a broad ensemble of global and regional models that could be used. Each combination of such models encapsulates different modelling assumptions and therefore would provide a distinct estimate of the behaviour of maximum temperatures over Ireland. To fully quantify the uncertainty in our estimates would require an adequate sampling strategy to select global-regional model combinations from the available ensemble. 

Our analysis looks only at the change in extreme temperature events over the historical record. We do not have observations of the future. Climate models are our only reliable tool for predicting future temporally non-stationary extremal behaviour under different scenarios, so being able to link the temporal non-stationarity of observational and climate data is essential to understand model strengths and weaknesses. This is particularly important for the consideration of highly nonlinear change linked to instabilities such as the possible effects of any change in the strength of the Atlantic meridional overturning circulation (AMOC) which has a profound modulating effect on Irish climate. Most global models suggest some weakening of the AMOC through to 2100, and a complete shutdown cannot be ruled out \citep{ipcc2021ch11}.

Finally, although we focus on temperature, many other variables (e.g., wind, rainfall) are important for assessing climate change for extreme meteorological events marginally, jointly and integrated over different time windows. Climate model data are likely to provide improvements in extremal inference of such joint distributions under the assumption that they better capture the physical interactions between processes. This can be used to enhance the equivalent empirical information from the observational data.\\

\subsection{Choice of threshold in non-stationary analysis}\label{const_thresh_discussion}
\noindent Threshold choice for the GPD and other tail models for identically distributed univariate extremes has been a major area of research for much of the last 40 years. Therefore, it is not surprising that there are a number of different perspectives for picking a systematic threshold selection criteria in our temporally non-stationary spatial context.

For univariate temporally non-stationary problems \cite{Eastoe2009} propose pre-processing the data using models fitted to the body of the distribution before modelling the extremes of the residuals with a constant threshold. Another approach is to use a conditional quantile \citep{Northrop2011}. As noted in Section~\ref{sec:above} it is difficult to account for the uncertainty in the threshold selection, so incorporating a temporal trend into the threshold undermines our ability to account for the uncertainty in estimating the temporal change in extreme data, which is our primary focus. We are pleased to see that even with our constant threshold at each station we found a simple model for how the GPD scale parameter changes over time and that the GPD is a good fit globally. More generally, the signal-to-noise ratio is critical in determining whether non-stationarity is accounted for in selecting the set of ``extreme data" to analyse.

We had an additional threshold to select for the extreme spatial dependence modelling via the risk function $r$. We had to face issues of missing data, with our approach presenting the first methods, we are aware of, for this. Climate models may help here either through exploring the sensitivity of different missing data patterns or through the use of reanalyses (weather forecast models conditioned on the observed values) to replace missing data, as these will help identify the largest events over space correctly.\\

\subsection{Choice of spatial extremes dependence model}
\noindent The scale of Ireland relative to the physical systems that drive temperature extremes has also played a key role in our choice of extreme value approach for modelling spatial dependence. This enabled us to take a simple model which is asymptotically dependent at the largest required spatial separation, which we achieved via an $r$-Pareto process coupled to a log-Gaussian latent process with a bounded variogram. We do not believe our approach would be applicable at much broader scales.

Even over the scale of Ireland, the asymptotic dependence property will not necessarily hold for other climatic variables, e.g., precipitation, which are manifest on smaller scales and with higher variability. In such cases, the modelling approaches need to incorporate asymptotic independence and to address issues about the scale over which asymptotic dependence holds \citep{Wadsworth2019} or even whether the spatial process is stationary over different mixture type events, e.g., convective or frontal precipitation \citep{Richards2023}. Any application of this method to different regions or processes should certainly involve an assessment of the evidence for asymptotic dependence, as our impression is that this assumption is made too readily. We find that, as a good approximation, we can assume spatial stationarity and isotropy. Ireland has, relatively speaking, simple topography with few ranges of hills of significant altitude. It does not follow that the method would be readily applicable to more complex alpine regions without, at a minimum, considerable additional validation.\\

\subsection{Choice of metrics for assessing change}
\noindent Section~\ref{SEC5} illustrates the challenge of finding effective metrics to illustrate temporal change when considering extremes of spatial fields. Metrics for marginal variables, e.g., high quantiles, are well-established and parsimonious. We see the development of spatial risk measures which enable the simple assessment of changing risk over time as an important avenue for further research. Spatial extreme value model inference also lacks well-established diagnostic methods for assessing the fit with observed events. Pairwise, and potentially higher order versions of the measure $\chi$, used in Sections~\ref{SEC4_2} and \ref{SEC5}, can be helpful but these may not be sufficient in practice. \cite{Winter2016} used severity-area frequency curves as a basis of comparison, but these focused only on assessing the performance of the model for the dependence structure. Picking metrics that directly link to risk assessment from heatwaves, such as health factors \citep{Winter2016} or crop failures or forest fires \citep{Zhang2022}, is likely to be valuable for planners. 

We focused on the spatial properties of the extreme events. Models for spatio-temporal extremal dependence of the process are needed to capture the evolution over time of spatial extreme events. This is an area where greater focus is required. For processes that are asymptotically dependent in space and time, some methods have been developed \citep{Davis2013a,Huser2014}, and it is pleasing to see recent extensions to incorporate asymptotic independence \citep{Simpson2021b}.\\

\section*{Acknowledgements}
\noindent Healy's work was supported by SFI grant 18/CRT/6049. Parnell’s work was supported by the SFI awards 17/CDA/4695; 16/IA/4520; 12/RC/2289\_P2 and we thank Simon Noone (Maynooth University) for help with data. For access to climate data, we acknowledge the World Climate Research Programme's Working Groups on Regional Climate and on Coupled Modelling, and the European Network for Earth System Modelling. We would like to thank all three referees who provided helpful and extensive constructive criticisms of the work and which has enormously improved its quality and presentation. \\
\bibliography{references}
\end{document}